\documentclass[twocolumn,showpacs,preprintnumbers,amsmath,amssymb]{revtex4} \usepackage{dcolumn}

\usepackage{graphicx} \usepackage{bm}

\newcommand{\hH}{\hat{H}}
\newcommand{\hvz}{\hat{v}_z}
\newcommand{\hvx}{\hat{v}_x}
\newcommand{\hvy}{\hat{v}_y}
\newcommand{\hpz}{\hat{p}_z}

\newcommand{\ha}{\hat{a}}

\newcommand{\hap}{\hat{a}^+}

\newcommand{\hB}{\hat{\cal B}}
\newcommand{\hBp}{\hat{\cal B}^+}

\newcommand{\hC}{\hat{\cal C}}
\newcommand{\hCp}{\hat{\cal C}^+}

\newcommand{\hJ}{\hat{\cal J}}
\newcommand{\hJp}{\hat{\cal J}^+}

\newcommand{\hM}{\hat{\cal M}}
\newcommand{\hO}{\hat{\Omega}}
\newcommand{\hcO}{\hat{\cal O}}
\newcommand{\hP}{\hat{\cal P}}

\newcommand{\hT}{\hat{\cal T}}

\begin{document}
\title{Zitterbewegung of Klein-Gordon particles and its simulation by classical systems}
\date{\today}
\author{Tomasz M. Rusin$^1$}
\email{Tomasz.Rusin@orange.com}
\author{Wlodek Zawadzki$^2$}
\affiliation{$^1$Orange Customer Service sp. z o. o., ul. Twarda 18, 00-105 Warsaw, Poland\\
             $^2$ Institute of Physics, Polish Academy of Sciences, Al. Lotnik\'ow 32/46, 02-688 Warsaw, Poland}

\begin{abstract}
The Klein-Gordon equation is used to calculate the Zitterbewegung
(ZB, trembling motion) of spin-zero particles in absence of fields and in the
presence of an external magnetic field. Both Hamiltonian and wave formalisms are
employed to describe ZB and their results are compared. It is demonstrated
that, if one uses wave packets to represent particles, the ZB motion has a decaying
behavior. It is also shown that the trembling motion is caused by an interference
of two sub-packets composed of positive and negative energy states which propagate
with different velocities. In the presence of a magnetic field the quantization of
energy spectrum results in many interband frequencies contributing to ZB oscillations
and the motion follows a collapse-revival pattern. In the limit of non-relativistic
velocities the interband ZB components vanish and the motion is reduced to
cyclotron oscillations. The exact dynamics of a charged Klein-Gordon particle
in the presence of a magnetic field is described on an operator level. The trembling
motion of a KG particle in absence of fields is simulated using a classical model
proposed by Morse and Feshbach -- it is shown that a variance of a Gaussian wave packet
exhibits ZB oscillations.
\end{abstract}

\maketitle

\section{Introduction \label{Sec_Intro}}

The phenomenon of Zitterbewegung (ZB, trembling motion) goes back to Schrodinger who proposed
it in 1930 for free relativistic electrons in a vacuum~\cite{Schrodinger1930}. Schrodinger
observed that, due to non-commutativity of the velocity operators with the Dirac Hamiltonian,
relativistic electrons experience a trembling motion in absence of external fields.
It was later recognized that ZB is due to an interference of electron states with positive and
negative electron energies. A very high frequency of ZB in a vacuum,
corresponding to~$\hbar\omega_Z=2m_ec^2$, and its very
small amplitude on the order of the Compton
wavelength~$\lambda_c=\hbar/m_ec\simeq 3.86\times 10^{-3}$~\AA\ made it impossible to observe
this effect in its original form with the currently available experimental methods.
However, in a recent
work Gerritsma {\it et al.}~\cite{Gerritsma2010} simulated the 1+1 Dirac equation and the
resulting Zitterbewegung with the use of trapped ions excited by laser beams. The important
advantage of this method is that one can simulate also the basic parameters of the Dirac equation
and tailor their desired values. The result of Gerritsma {\it et al.} allows one to expect that
observable effects for relativistic particles in a vacuum can be convincingly
reproduced with more ``user friendly'' parameters.
In general, there has been recently a revival of interest in the relativistic-type
equations related to ``the rise of graphene''~\cite{Novoselov2004}, topological
insulators and similar systems in narrow-gap semiconductors~\cite{Zawadzki2011}.

The purpose of our paper is to describe the phenomenon of Zitterbewegung for charged
Klein-Gordon (KG) spin-zero particles in absence of fields and in the presence of a magnetic
field~\cite{Klein1926,Gordon1926,Fock1926}.
The Zitterbewegung of KG particles in absence of fields was described before,
see~\cite{Fuda1982,WachterBook,GreinerBook}. However, in our treatment we introduce a number
of additional elements. First, we describe the particles by wave packets and show that
this feature leads to a transient character of the resulting ZB motion.
Second, we use both the Hamiltonian
and wave forms of Klein-Gordon equation (KGE) and show the equivalence of the two approaches.
Third, we point out that ZB is a result of interference
between positive and negative energy sub-packets
propagating with different velocities. Fourth, we simulate classically the ZB motion using a
simple mechanical system proposed by Morse and Feshbach~\cite{MorseBook}. Still, our main
objective is to consider in detail the dynamics of a charged KG particle in the presence of
an external uniform magnetic field and describe the phenomenon of ZB in this situation.
To the best of our knowledge this problem has not been treated before.

The one-particle Klein-Gordon equation for spin-zero particles leads to some well known
difficulties~\cite{WachterBook,SchweberBook}.
The KG equation involves second time derivative, the probability density
is not positively definite, there are problems with the position operator or vanishing
square of the velocity operator. For this reason in the present work we calculate
ZB of {\it average current} which has well defined meaning in the theory of KG equation.
For charged particles the average current is proportional to average particle velocity, so
in our work we calculate one of these two quantities. In previous treatments of ZB
for Dirac equation, simulation by trapped ions or solid-state systems,
the authors usually calculated ZB of the position operator.

In our considerations we encounter another interesting anomaly of KG equation, namely,
that particle velocities can exceed the speed of light for sufficiently large momenta.
In other words it appears that, in contrast to the Dirac equation for electrons,
KGE does not posses an automatic ``safety brake'' for velocities to keep them below~$c$.
To our knowledge this feature has not been remarked before, so we mention it throughout
our work.

Our paper is organized as follows. In Section~\ref{Sec_Vac} we calculate ZB of a
wave packet using the Hamiltonian formalism, in Section~\ref{Sec_Wave} we obtain
similar results with the use of KG waves and discuss explicitly physical background
for the transient behavior of ZB motion. Section~\ref{Sec_Field} contains a description
of ZB for a charged KG particle in a magnetic field. In Section~\ref{Sec_Sim} we simulate
classically the ZB phenomenon using a system proposed by Morse and Feshbach. In Section~\ref{Sec_Disc}
we discuss our results, the paper is concluded by a summary. Appendix~\ref{App_BJ} contains a
derivation of particle dynamics in the presence of a magnetic field, Appendix~\ref{App_Light}
discusses the problem of high particle velocities, in Appendices~\ref{App_Ident} and~\ref{App_Gauss}
we give some mathematical details.

\section{Zitterbewegung in vacuum \label{Sec_Vac}}

We begin by considering a Klein-Gordon particle in absence of external fields.
The Klein-Gordon equation in the Hamiltonian form is~\cite{Feshbach1958}
\begin{equation} \label{H_KG}
i\hbar \frac{\partial \Psi}{\partial t}= \hH\Psi.
\end{equation}
Here the Hamiltonian is
\begin{equation} \label{H_H}
 \hH = \frac{\tau_3+i\tau_2}{2m}\hat{\bm p}^2+\tau_3mc^2,
\end{equation}
where~$m$ is particle mass,~$\hat{\bm p}$ is particle momentum
and~$\tau_j$~($j=1,2,3$) are the Pauli matrices~$\sigma_j$, respectively.
The wave function~$\Psi$ is a two-component vector
\begin{equation} \label{H_Psi}
 \Psi = \left( \begin{array}{c} \varphi \\ \chi \end{array}\right).
\end{equation}
In the Hamiltonian form one can introduce the Heisenberg picture~\cite{Feshbach1958}.
The~$z$-th component of the time-dependent velocity operator is
\begin{equation} \label{H_vz}
 \hvz(t) = e^{i\hH t/\hbar} \hvz(0) e^{-i\hH t/\hbar},
\end{equation}
where~$\hvz(0)= \partial \hH /\partial \hpz$. In this representation~$\hvz(t)$ is a~$2\times 2$
matrix operator. Expanding~$e^{i\hH t/\hbar}=1+\hH t + (1/2!)\hH^2+\ldots$ and noting
that~$\hH^2=E^2$, where the energy is~$E=\pm cp_0$ with
\begin{equation} \label{H_p0}
p_0 = +\sqrt{m^2c^2+p^2},
\end{equation}
we obtain
\begin{equation} \label{H_eiHt}
 e^{i\hH t/\hbar} = \cos(Et/\hbar) + \frac{i\hH}{E}\sin(Et/\hbar).
\end{equation}
The velocity operator in Eq.~(\ref{H_vz}) is a product of three matrices. Its~$(1,1)$ component is
\begin{equation} \label{H_vz_11}
(\hvz)_{11}(t) = \frac{\hpz}{m} + \frac{\hat{p}^2\hpz}{2m\hat{p}_0^2}\left[\cos(2Et/\hbar) -1 \right].
\end{equation}
The remaining elements of~$\hvz(t)$ are calculated similarly. The~$\hat{v}_x$
and~$\hat{v}_y$ components of the velocity operator are obtained from~$\hvz(t)$ by the
replacement~$\hpz \rightarrow \hat{p}_x, \hat{p}_y$, respectively.
In the non-relativistic limit~$p \ll mc$ we obtain in Eq.~(\ref{H_vz_11}) the classical
motion~$(\hvz)_{11}(t) \simeq \hpz/m$.
In absence of external fields~$p_i$ are good quantum numbers. We introduce~${\bm p}=\hbar{\bm k}$
and~${\bm q}=\lambda_c{\bm k}$, where the effective Compton wavelength is~$\lambda_c=\hbar/mc$.
Also, we introduce a useful frequency~$\omega_0=(mc^2)/\hbar$. Both~$\lambda_c$ and~$\omega_0$
refer to particles of mass~$m$. In the above notation Eq.~(\ref{H_vz_11}) becomes
\begin{equation} \label{H_vz_11q}
(\hvz)_{11}(t) = cq_z + \frac{c}{2}\frac{q^2q_z}{(1+q^2)}\left[\cos(2\omega_0 t\sqrt{1+q^2}) -1 \right].
\end{equation}

The first term in Eq.~(\ref{H_vz_11q}) corresponds to the classical motion of a particle
while the second term describes rapid oscillations of the velocity.
The velocity oscillates from~$v_{max}=cq_z$ to~$v_{min}=cq_z/(1+q^2)$.
Since the maximum velocity of the particle is~$c$, there must be~$|{\bm q}|\le 1$.
We notice that, in principle, Eq.~(\ref{H_vz_11q}) admits velocities above the
speed of light. We discuss this issue in more detail in Appendix~\ref{App_Light}.
The frequency of oscillations varies from~$\omega=2\omega_0$ for low~${\bm q}$
to~$\omega=2\sqrt{2}\omega_0$ for~$|\bm q|=1$. The velocity oscillations taking place
in absence of external fields are called Zitterbewegung.

Integrating~$(\hvz)_{11}(t)$ in Eq.~(\ref{H_vz_11q}) over time we have
\begin{eqnarray} \label{H_rz_11q}
\hat{z}_{11}(t) &=& z_{11}(0) + cq_zt - \frac{c}{2}\frac{q^2q_z}{1+q^2}t + \nonumber \\
&&\frac{\lambda_c}{4}\frac{q^2q_z}{(1+q^2)^{3/2}}\sin(2\omega_0 t\sqrt{1+q^2}).
\end{eqnarray}
The amplitude of ZB oscillations of the position operator is on the order of~$\lambda_c$.
The operator~$\hat{z}_{11}(t)$ is obtained in the formal way, physical
limitations to the position operator will be discussed below.

In order to obtain physical observables one needs to average the operator quantities
over the wave packet.
The average velocity~$\langle \hvz(t)\rangle$ of the wave packet~$|W\rangle$ is
\begin{equation} \label{P_vz}
\langle \hvz(t)\rangle = \langle W|\tau_3\hvz(t)|W\rangle =
 \sum_{{\bm p}{\bm p}'} \langle W|{\bm p}\rangle \langle {\bm p}|\tau_3\hvz(t)|{\bm p}'\rangle
 \langle {\bm p}'|W\rangle.
\end{equation}
For KGE in the Hamiltonian form the matrix elements of operators include
an additional~$\tau_3$ factor~\cite{Feshbach1958}.
We take the wave packet in the form of a two-component
vector~$\langle {\bm r}|W\rangle=(1,0)^T \langle {\bm r}|w\rangle$ with one non-vanishing
component. Here~$\langle {\bm r}|w\rangle \equiv w(\bm {r})$ is a
Gaussian function with a nonzero momentum~$\hbar {\bm k}_0$
\begin{equation} \label{P_wr}
 w(\bm {r})=\frac{1}{(d\sqrt{\pi})^{3/2}} \exp[-r^2/(2d^2) + i{\bm k_0}{\bm r}].
\end{equation}
There is~$w(\bm {k})=\int e^{-i{\bm k}{\bm r}/\hbar} w({\bm r})d^3{\bm r}$ and we have
\begin{equation} \label{P_wk}
 \langle {\bm k}|w\rangle=(2d\sqrt{\pi})^{3/2}\exp[-d^2({\bm k}-{\bm k}_0)^2/2].
\end{equation}
The wave packet~$|W\rangle$ selects~$(1,1)$ component of the velocity matrix~$\hvz(t)$.
From Eqs.~(\ref{P_vz}) and~(\ref{P_wk}) we obtain
\begin{eqnarray} \label{P_vzq}
\langle \hvz(t)\rangle &=& c\frac{{d_c^3}}{\pi^{3/2}}
 \int \exp[-d_c^2({\bm q}-{\bm q}_0)^2] \nonumber \\ && \left\{ q_z + \frac{1}{2}\frac{q^2q_z}{1+q^2}
    \left[\cos(\omega_0 t\sqrt{1+q^2})-1\right] \right\} d^3{\bm q}, \ \ \ \ \ \
\end{eqnarray}
where~$d_c=d/\lambda_c$. This integral is nonzero only if~${\bm q}_0$ has a
nonzero~$z$-th component, so we take~${\bm q}_0=(0,0,q_{0z})^T$.
Selecting the~$z$ axis to be parallel to~${\bm q}_0$ and using the spherical coordinates
we calculate the integrals over the two angular variables.
The remaining integral over~$q$ is computed numerically.

\begin{figure}
\includegraphics[width=8cm,height=8cm]{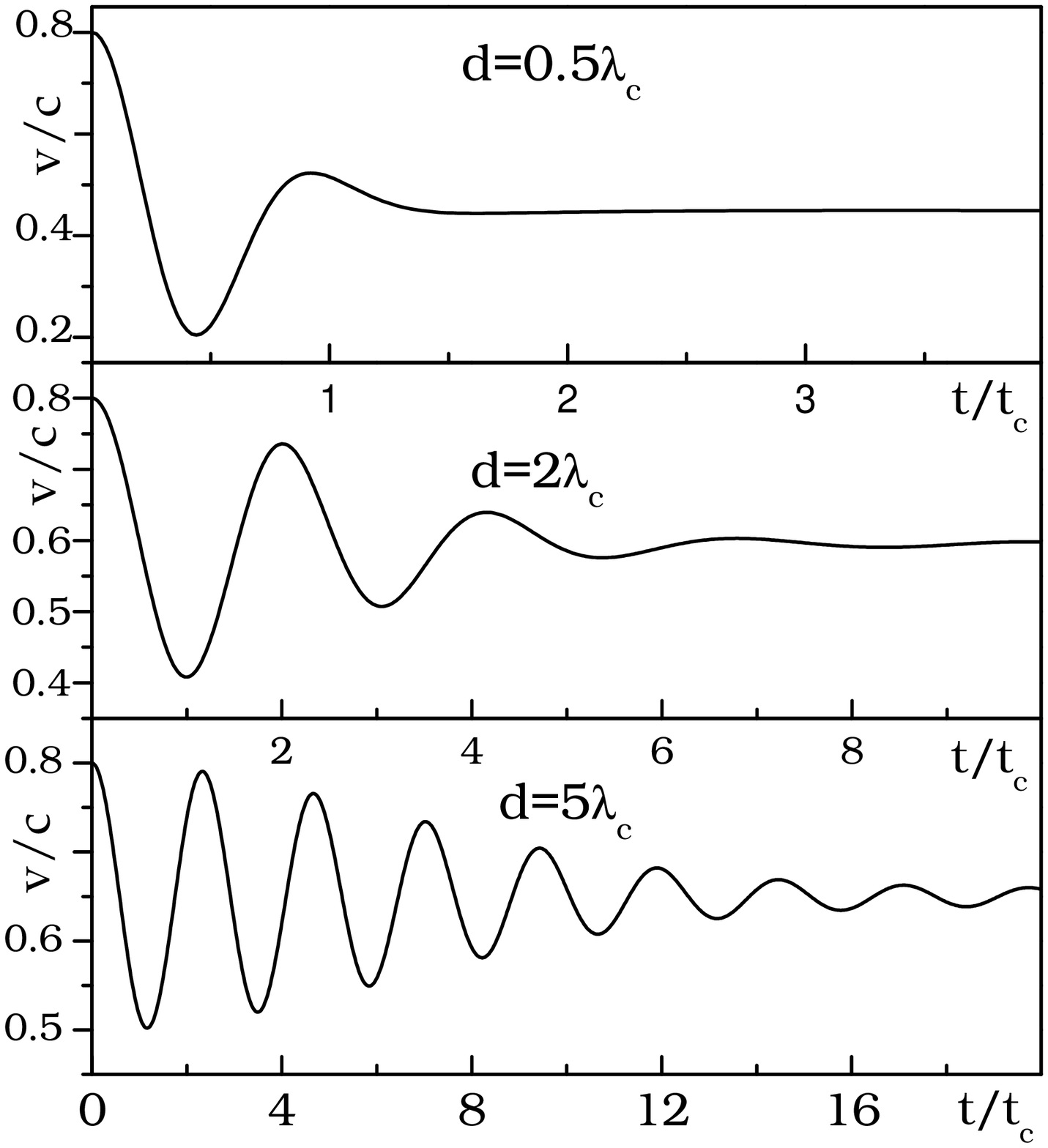}
\caption{\label{Figure1} Calculated velocity of wave packet in absence of external
         fields for three packet widths~$d$.
         The phenomenon of transient Zitterbewegung is seen.
         The initial packet wave vector is~${\bm k}_0=(0,0,k_{0z})$
         with~$k_{0z}=0.8\lambda_c^{-1}$. Time is expressed in~$t_c=\hbar/mc^2$ units.
         Initial packet velocity is~$v_{0z}=\hbar k_{0z}/m =0.8c$,
         its final velocity depends on packet parameters.}
\end{figure}

In Fig.~\ref{Figure1} we plot the average packet velocity~$\langle \hvz(t)\rangle$ calculated
from Eq.~(\ref{P_vzq}) for three different packet widths~$d$.
The time on the horizontal axis is expressed
in~$t_c=\hbar/mc^2$ units, where~$t_c=(m_e/m) \times 1.29\times 10^{-21}$~s and~$m_e$ is
the electron mass. In all cases the motion has a transient character.
Physically, the decay of ZB oscillations is due to different propagation
velocities of sub-packets corresponding to the positive and negative energy states.
We analyze this effect below. It is seen that the final packet velocity
differs from the initial value~$\hbar k_{0z}/m$. In the limit of~$d\rightarrow \infty$
the velocity oscillations do not decay in time.

Now we calculate the average velocity by splitting the initial wave packet into
two sub-packets corresponding to the positive and negative energy states.
First we introduce the unity operator~\cite{Fuda1982}
\begin{equation} \label{Av_1}
\hat{1} = \sum_{{\bm k}s} |{\bm k}s\rangle\langle {\bm k}s|\tau_3s,
\end{equation}
where~$s=\pm 1$, and
\begin{equation} \label{Av_kr}
\langle {\bm r} |{\bm k}s\rangle =
 \frac{e^{i{\bm k}{\bm r}}}{2\sqrt{mcp_0}}
 \left(\begin{array}{c} mc + s p_0 \\ mc - s p_0 \end{array}\right)
\end{equation}
are the two eigenstates of~$\hH$ corresponding to the positive and negative
energies~$E_{s}=s cp_0$. These states are normalized
according to~$\langle {\bm k}s|\tau_3|{\bm k}'s'\rangle/(2\pi)^{3/2} =
s \delta_{{\bm k}{\bm k}'}\delta_{ss'}$. Then
\begin{equation} \label{Av_W}
|W\rangle = \sum_{{\bm k}s}s |{\bm k}s\rangle \langle {\bm k}s|\tau_3|W\rangle
 = \sum_{{\bm k}s}s |{\bm k}s\rangle W_{{\bm k}s},
\end{equation}
where~$W_{{\bm k}s}=\langle {\bm k}s|\tau_3|W\rangle$. The sub-packet of positive
energy states is~$|W+\rangle=\sum_{\bm k}|{\bm k}+\rangle W_{{\bm k}+}$,
while the sub-packet of negative energy states is~$|W-\rangle=\sum_{\bm k}|{\bm k}-\rangle W_{{\bm k}-}$.
Using Eqs.~(\ref{Av_kr}) and~(\ref{Av_W}) we find
\begin{equation} \label{Av_Wks}
 W_{{\bm k}s}= (2d\sqrt{\pi})^{3/2}\frac{(mc+s p_0)}{2\sqrt{mcp_0}}
    e^{-d^2({\bm k}-{\bm k}_0)^2/2}.
\end{equation}
The average packet velocity is
\begin{eqnarray} \label{Av_vz}
\langle \hvz(t)\rangle &=&
     \sum_{{\bm k}{\bm k}'ss'}s s' W_{{\bm k}s}^* W_{{\bm k}'s'}
     \langle {\bm k}s|\tau_3\hvz(t)|{\bm k}'s' \rangle
     \nonumber \\
&=&\sum_{{\bm k}{\bm k}'ss'}s s' W_{{\bm k}s}^* W_{{\bm k'}s'}
e^{i(\omega_{s}-\omega_{s'})t}
\langle {\bm k}s|\tau_3 \frac{\partial \hH}{\partial \hat{p_z}}|{\bm k}'s'\rangle. \ \ \ \ \ \
\end{eqnarray}
We defined~$\omega_{s}=s\omega_0\sqrt{1+(k\lambda_c)^2}$ and used the equality
\begin{equation} \label{Av_tauH}
\langle {\bm k}s|\tau_3 e^{i\hH t/\hbar}=\langle {\bm k}s|e^{i\hH^{\dagger} t/\hbar}\tau_3
=e^{i\omega_{s}t}\langle {\bm k}s|\tau_3,
\end{equation}
which follows from the properties:~$\hH=\tau_3\hH^{\dagger}\tau_3$
and~$\langle {\bm k}s|\hH^{\dagger}=(\hH |{\bm k}s\rangle)^{\dagger}
=E_{s}\langle {\bm k}s|$. Another proof of the identity~(\ref{Av_tauH}) is given in
Appendix~\ref{App_Ident}. There is also
\begin{equation} \label{Av_vkk}
\langle {\bm k}s|\tau_3 \frac{\partial \hH}{\partial p_z}|{\bm k}'s'\rangle
   = (2\pi)^3\frac{cp_z}{p_0}\delta_{{\bm k}{\bm k}'},
\end{equation}
which does not depend on~$s$ and~$s'$.
Combining Eqs.~(\ref{Av_vz}) -~(\ref{Av_vkk}) we obtain
\begin{eqnarray} \label{Av_vzk}
\langle \hvz(t)\rangle &=& \frac{2d^3\pi^{3/2}}{(2\pi)^3m} \int
    \frac{p_z}{p_0^2}e^{-d^2({\bm k}-{\bm k}_0)^2} d^3{\bm k} \nonumber \\
 & \times &\sum_{s,s'}ss'(mc+s p_0)(mc+s' p_0)
     e^{i(\omega_{s}-\omega_{s'})t}. \ \ \ \ \
\end{eqnarray}
The average velocity in Eq.~(\ref{Av_vzk}) is a sum of four terms. The term with~$s=s'=+1$
describes the motion of positive energy sub-packet,
while the term with~$s=s'=-1$ corresponds to the negative energy sub-packet
\begin{equation} \label{Av_vz_pm}
\langle\hat{v}_z\rangle^{\pm}= \frac{d^3}{4m\pi^{3/2}} \int
  \left(1\pm \frac{mc}{p_0}\right)^2 p_z e^{-d^2({\bm k}-{\bm k}_0)^2} d^3{\bm k}.
\end{equation}
Thus the two sub-packets move with different velocities. Their
relative velocity is
\begin{equation} \label{Av_vz_rel}
\langle\hat{v}_z\rangle^{rel}= \frac{cd^3}{\pi^{3/2}} \int \frac{p_z}{p_0}
e^{-d^2({\bm k}-{\bm k}_0)^2} d^3{\bm k}.
\end{equation}
Two terms in Eq.~(\ref{Av_vzk}) with~$s\neq s'$, corresponding to an interference of
the two packets, give rise to an oscillatory term
\begin{eqnarray} \label{Av_vz_osc}
\langle\hat{v}_z(t)\rangle^{osc}&=& \frac{d^3}{4m\pi^{3/2}} \int
    \left(1-\frac{m^2c^2}{p_0^2}\right) p_z \nonumber \\
    &\times & \cos(2\omega_kt) e^{-d^2({\bm k}-{\bm k}_0)^2} d^3{\bm k},
\end{eqnarray}
where~$\omega_k=\sqrt{1+(k\lambda_c)^2}$.
According to the Riemann-Lesbegues theorem this term has a transient character~\cite{Lock1979}.
Performing integrations in Eqs.~(\ref{Av_vz_pm}) and~(\ref{Av_vz_osc}),
we obtain again Eq.~(\ref{P_vzq}).
Thus we showed that the ZB oscillations arise from the interference of
positive and negative energy states. After a certain time the two
sub-packets are sufficiently far away from each other and the overlap between
them vanishes, which results in the disappearance of ZB oscillations.
This explains the behavior of velocity shown in Fig.~\ref{Figure1}.

To evaluate the decay of ZB oscillations, we estimate the
time after which the two sub-packets will be separated from each other by the distance~$2d$.
Assuming that~$k_0\lambda_c \simeq 1$, the relative velocity between the two sub-packets
is~$\langle\hat{v}_z\rangle^{rel} \simeq c (k_0\lambda_c)$.
The time interval after which the distance between the sub-packets exceeds~$2d$ is
\begin{equation} \label{Av_td}
 t_d \simeq \frac{2d}{ck_0\lambda_c}.
\end{equation}
It is seen in Fig.~\ref{Figure1} that the ZB oscillations nearly
disappear after~$t_d$. For example there is~$t_d=5t_c$ for~$d=2\lambda_c$.
Since the ZB frequency is~$2\omega_0=2mc^2/\hbar$,
a number of non-vanishing oscillations is approximately
\begin{equation} \label{Av_Nosc}
 N_{osc} \simeq \frac{2\omega_0 t_d}{2\pi} = \frac{2}{\pi}\left(\frac{d}{\lambda_c}\right)
  \left(\frac{1}{k_0\lambda_c}\right).
\end{equation}
The above estimation correctly evaluates the number of ZB oscillations seen
in Fig.~\ref{Figure1}. The optimal conditions for an appearance of ZB are: wide packets and
small values of~$|{\bm k}_0|$. On the other hand, for too small values of~$|{\bm k}_0|$
one of the two sub-packets disappears, see Eq.~(\ref{Av_vz_pm}),
which reduces amplitude of ZB oscillations.

\section{Wave form of KGE \label{Sec_Wave}}
Now we intend to demonstrate a relation between the ZB oscillations of the average packet velocity
calculated above with the use of the Hamiltonian form of KGE and
an average current obtained from the wave form of KGE. In absence of external
fields the Klein-Gordon equation has the wave equation form
\begin{equation} \label{WE_KG}
\frac{1}{c^2}\frac{\partial^2}{\partial t^2} \phi(x) -{\bm\nabla}^2\phi(x)
+ \frac{ m^2c^2}{\hbar^2}\phi(x) =0,
\end{equation}
where~$x=(ct,{\bm r})$ is the position four-vector~\cite{WachterBook}.
The solution of this equation is
\begin{equation} \label{WE_phi}
\phi(x)= \frac{1}{(2\pi)^{3}} \int \sqrt{\frac{mc}{p_0}}
    \left[a({\bm k})e^{-ik\cdot x} + b^*({\bm k})e^{ik \cdot x} \right]d^3{\bm k},
\end{equation}
where~$k=(\omega_k/c,{\bm k})$,~$\omega_k=\omega_0\sqrt{1+(k\lambda_c)^2}$,
and~$a({\bm k})$,~$b^*({\bm k})$ are complex coefficients.
Function~$\phi$ is normalized to
\begin{equation} \label{WE_Norma}
\frac{i\hbar}{2mc^2}\int \left[\phi^*\frac{\partial \psi}{\partial t} -
         \left(\frac{\partial \phi^*}{\partial t} \right)\psi \right]d^3{\bm r} = Q,
\end{equation}
where~$Q=\pm 1$ for charged particles and~$Q=0$ for neutral particles.
In the following we select~$Q=+1$, which leads to
\begin{equation} \label{WE_ab}
\int d^3{\bm k} \left[a^*({\bm k})a({\bm k}) - b^*({\bm k})b({\bm k}) \right] = 1.
\end{equation}
To determine the coefficients~$a({\bm k})$ and~$b^*({\bm k})$ we need two boundary
conditions for~$\phi$ and~$\partial\phi/\partial t$ at~$x=(0,{\bm r})$.
Having specified~$a({\bm k})$ and~$b^*({\bm k})$
one can calculate the current density~${\bm j}(x)$
\begin{equation} \label{WE_j}
{\bm j}(x) = \frac{\hbar}{2im}\left[\phi^*({\bm \nabla}\phi) - ({\bm \nabla}\phi^*)\phi \right],
\end{equation}
and the average current~$\langle {\bm j}(t)\rangle = \int {\bm j}(x) d^3{\bm r}$.

Our aim is to find a correspondence between the average packet velocity calculated
in Eq.~(\ref{P_vzq}) and the average current~$\langle {\bm j}(t)\rangle$
given in Eq.~(\ref{WE_j}). To this end we select the coefficients~$a({\bm k})$
and~$b^*({\bm k})$ in such a way
that the function~$\phi$ in the wave form of KGE corresponds to the wave
packet~$(w(\bm r),0)^T$ in the Hamiltonian form of KGE.
Relations between~$\phi$,~$\partial \phi/\partial t$ and the two-component wave
function~$\Psi=(\varphi,\chi)^T$ in the Hamiltonian form of KGE are~\cite{WachterBook}
\begin{eqnarray}
\phi &=&\varphi + \chi \label{WE_phi0}, \\
i\partial \phi/\partial t &=& mc^2(\varphi - \chi)/\hbar. \label{WE_phi1}
\end{eqnarray}
Since~$(\varphi,\chi)^T=(w(\bm r),0)^T$ we find the coefficients~$a({\bm k})$ and~$b^*({\bm k})$
from Eqs.~(\ref{WE_phi0}) -~(\ref{WE_phi1}) by setting~$\varphi(t=0,{\bm r}) = w({\bm r})$ and~$\chi=0$.
From Eq.~(\ref{WE_phi0}) we have
\begin{eqnarray} \label{WE_bc0}
\int \sqrt{\frac{mc}{p_0}}
   \left[a  ({\bm k})e^{+i{\bm k}\cdot {\bm r}}
  +      b^*({\bm k})e^{-i{\bm k}\cdot {\bm r}} \right] d^3{\bm k} \nonumber \\
 = (2d\sqrt{\pi})^{3/2} \int e^{-({\bm k}-{\bm k}_0)^2d^2/2+i{\bm k}\cdot {\bm r}} d^3{\bm k},
\end{eqnarray}
while from Eq.~(\ref{WE_phi1}) we have
\begin{eqnarray} \label{WE_bc1}
\int \sqrt{\frac{mc}{p_0}}
   \left[ a  ({\bm k})e^{+i{\bm k}\cdot {\bm r}}
         -b^*({\bm k})e^{-i{\bm k}\cdot {\bm r}} \right] p_0 d^3{\bm k} \nonumber \\
 = (2d\sqrt{\pi})^{3/2}mc\int e^{-({\bm k}-{\bm k}_0)^2d^2/2+i{\bm k}\cdot {\bm r}} d^3{\bm k}.
\end{eqnarray}
In terms including~$b^*({\bm k})$ we replace~${\bm k} \rightarrow -{\bm k}$,
solve equations~(\ref{WE_bc0}) and~(\ref{WE_bc1})
for~$a({\bm k})$ and~$b^*(-{\bm k})$, and obtain
\begin{eqnarray} \label{WE_phi_ab}
\phi({\bm r},t) &=& \frac{(2d\sqrt{\pi})^{3/2}}{2(2\pi)^3} \int d^3{\bm k}
         e^{-d^2({\bm k}-{\bm k}_0)^2/2+i{\bm k}\cdot {\bm r}}
  \nonumber \\
  &\times& \left[\left(1 + \frac{mc}{p_0} \right)e^{-i\omega_kt}
             +\left(1 - \frac{mc}{p_0} \right)e^{+i\omega_kt} \right].
\end{eqnarray}
The above function~$\phi$ includes both positive and negative energy amplitudes.
For~$p\rightarrow 0$ there is~$1+mc/p_0 \simeq 2$ and~$1-(mc/p_0)\simeq p^2/2(mc)^2$.
Thus the second term in Eq.~(\ref{WE_phi_ab}) is much smaller than the first.
In this limit the packet consists of the positive energy states alone.

\begin{figure}
\includegraphics[width=8cm,height=8cm]{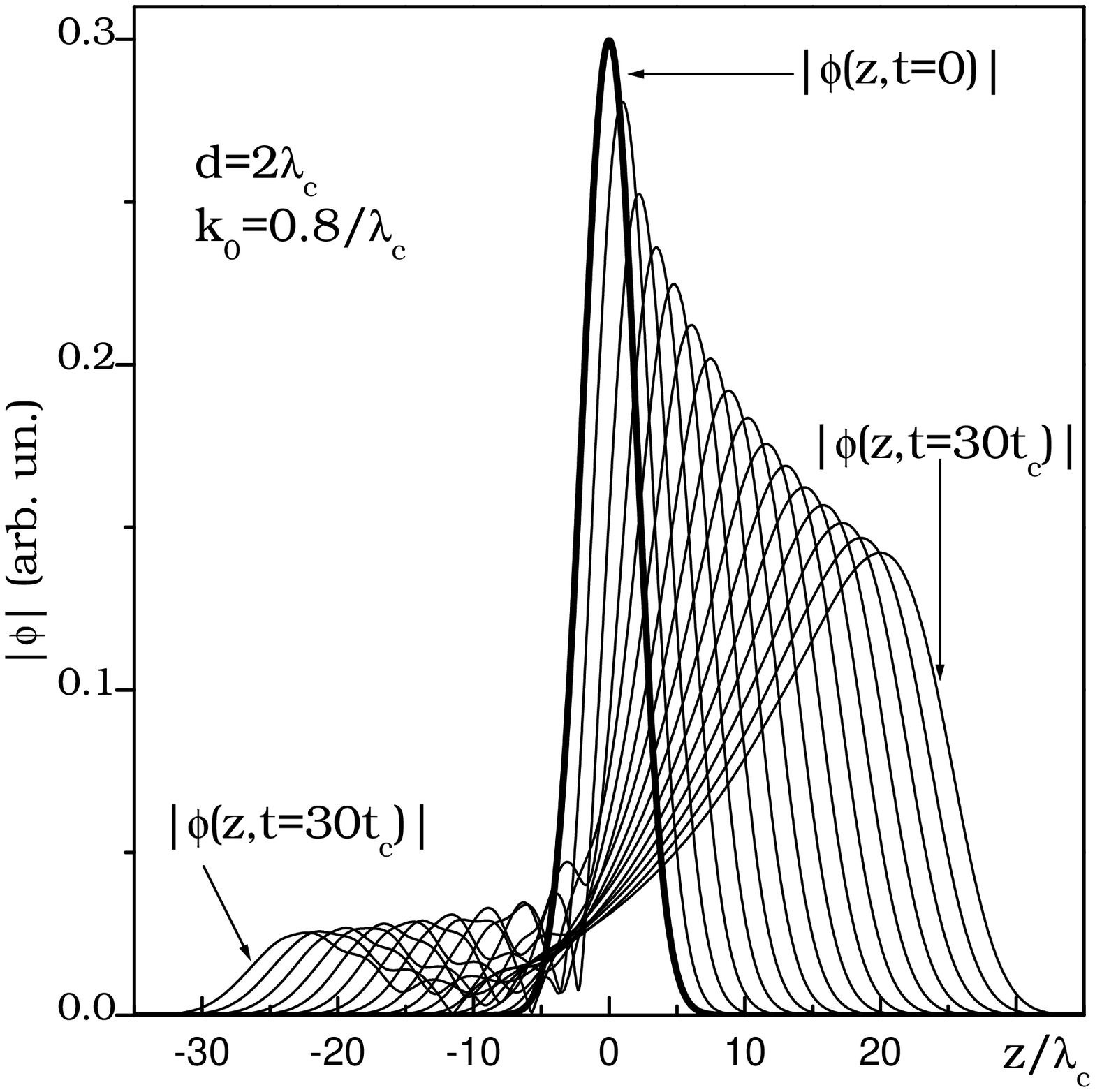}
\caption{\label{Figure2} Time evolution of the wave packet (absolute value) according to
         one-dimensional version of Eq.~(\ref{WE_phi_ab}). Initial wave packet (thick line)
         splits into two sub-packets moving with different velocities. Thin lines
         show shapes of sub-packets in successive time intervals~$2t_c=2\hbar/(mc^2)$.}
\end{figure}

In Fig.~\ref{Figure2} we plot the time evolution of the wave packet~$\phi$ in one dimension.
The packet propagates according to a one-dimensional version of Eq.~(\ref{WE_phi_ab}).
The initial packet is assumed in a Gaussian form
\begin{equation}
\phi(x,0)=\frac{d}{\sqrt{\pi}}e^{-x^2/(2d^2)+ik_0x}.
\end{equation}
Its absolute value is indicated in Fig.~\ref{Figure2} by the thick line.
Each thin line describes~$|\phi(x,t)|$
in successive time intervals~$2t_c=2\hbar/(mc^2)$. It is seen that the packet splits
into two sub-packets moving with different velocities. The sub-packet at the right
corresponds to positive energies while the sub-packet at the left corresponds
to negative energies. The difference in the amplitudes of sub-packets results
from different contributions of the positive and negative energy states in the initial packet
at~$t=0$, see Eqs.~(\ref{WE_bc0}) -~(\ref{WE_bc1}). The Zitterbewegung occurs only when
the sub-packets overlap. Each of the sub-packets slowly spreads in time,
but the spreading time is much larger than the overlapping time, so the
ZB vanishes much faster than the spreading of sub-packets.

Now we continue the calculation of average current given in Eq.~(\ref{WE_j})
using function~$\phi$ of Eq.~(\ref{WE_phi_ab}). This function has the form
of an integral over~${\bm k}$. To calculate the spatial derivative~$\nabla \phi$
we change the order of integration and differentiation, which can be done
for any function decaying exponentially for~$k\rightarrow \infty$.
Using the identity:~$1+(mc/p_0)^2=2 - (p/p_0)^2$, we obtain for the first term
of the average current
\begin{eqnarray} \label{WE_curr}
\frac{\hbar}{2im}\int \phi^*\frac{\partial \phi}{\partial z} d^3{\bm r} = \frac{d^3\hbar}{8im\pi^{3/2}}
\int d^3{\bm k} e^{-d^2({\bm k}-{\bm k}_0)^2}(ik_z) \times \nonumber \\
\left|\left(1 + \frac{mc}{p_0} \right)e^{-i\omega_kt}
              +\left(1 - \frac{mc}{p_0} \right)e^{+i\omega_kt} \right|^2
             \nonumber \\
 = \frac{d^3}{2\pi^{3/2}}\int\! d^3{\bm k} e^{-d^2({\bm k}-{\bm k}_0)^2}\!\!
 \left\{\frac{p_z}{m} +\frac{p_zp^2}{2mp_0}\left[\cos(2\omega_kt)-1 \right] \right\}. \ \
\end{eqnarray}
Calculation of the second term in the
current:~$\hbar/(2im)\int (\partial \phi^*/\partial z) \phi d^3{\bm r}$,
gives the same result but with an opposite sign,
so that both terms in Eq.~(\ref{WE_j}) add together.
Comparing Eq.~(\ref{WE_curr}) with Eqs.~(\ref{WE_j}) and~(\ref{P_vzq}) we conclude
that the current density~$\langle j_z(t)\rangle$ averaged over
the packet~$\phi(x)$ in Eq.~(\ref{WE_phi})
equals to the average velocity~$\langle v_z(t)\rangle$ of the packet
in the Hamiltonian form of KGE multiplied by the particle charge.
This way we establish an equivalence of Zitterbewegung in the Hamiltonian
and wave equation formalisms.

The above equivalence is valid for the average values only. In the Hamiltonian form
of KGE one can define the time dependent velocity operator
$\hat{\bm v}(t)=e^{i\hH t/\hbar} \hat{\bm v}(0)e^{-i\hH t/\hbar}$,
which can be expressed in a closed form without specifying of
the wave packet, see Eq.~(\ref{H_vz_11}). But an analogous current
operator in the wave form of KGE
can be defined as a current density~${\bm j}(x)$, which strongly depends on the form
of function~$\phi$.

Even more significant differences between the Hamiltonian and wave descriptions of ZB
appear in the analysis of the position operator~$\hat{\bm r}(t)$. In the Hamiltonian
form of KGE the position operator written in the Heisenberg
picture is~$\hat{\bm r}(t)=e^{i\hH t/\hbar} \hat{\bm r}(0)e^{-i\hH t/\hbar}$ and,
for the field-free KGE, it can be calculated in a compact form, see Eq.~(\ref{H_rz_11q})
and Ref.~\cite{Fuda1982}. On the other hand, there
is no well defined position operator~$\hat{\bm r}$ for the wave form of KGE since this
operator is not hermitian, see Ref.~\cite{SchweberBook}.
However, one can calculate an {\it average position operator} for the
wave form of KGE by integrating the average current over time
\begin{equation} \label{WE_rav}
\langle {\bm r}(t) \rangle = \langle {\bm r}(0) \rangle
        +\frac{1}{Q} \int \langle {\bm j}(t) \rangle dt,
\end{equation}
where the charge~$Q\neq 0$. This example indicates that the equivalence between
the Zitterbewegung for the Hamiltonian and wave equation formalisms holds
for the average values only.

\section{Zitterbewegung in a magnetic field \label{Sec_Field}}
In the presence of a magnetic field the KG Hamiltonian for a charged particle reads~\cite{WachterBook}
\begin{equation} \label{B_H}
 \hH = \frac{\tau_3+i\tau_2}{2m}(\hat{\bm p}-q{\bm A})^2+\tau_3mc^2,
\end{equation}
where~$q$ is the particle charge and~${\bm A}$ is the vector potential of a magnetic field.
We assume the magnetic field~${\bm B}$ to be parallel to the~$z$ axis
and describe it by the asymmetric gauge~${\bm A}=B(-y,0,0)$.
Eigenstates of the Hamiltonian are of the form
\begin{equation} \Psi({\bm r}) = e^{ik_xx+ik_zz}\Phi(y), \end{equation}
and the resulting eigenenergy equation is~$\hH \Psi = E \Psi$ with
\begin{equation}
 \hH =(\tau_3+i\tau_2) \frac{1}{2m}\left[(\hbar k_x+qBy)^2 + \hbar^2k_y^2+\hbar^2k_z^2\right]+\tau_3mc^2.
\end{equation}
We introduce the magnetic radius~$L = \sqrt{\hbar/|q|B}$ and
define~$\xi=k_xL + \eta_q y/L$, where~$\eta_q=\pm 1$ is the sign of~$q$.
Then there is~$\eta_q y=\xi L - k_xL^2$
and~$\partial/ \partial y = (1/L)\partial/ \partial \xi$.
The eigenenergies are~$ E_{\rm n} = s E_{n,k_z}$, where~\cite{RadovanovicBook}
\begin{equation} \label{B_E}
 E_{n,k_z} = \sqrt{m^2c^4 + 2mc^2\hbar \omega_c (n +1/2) + (c \hbar k_z)^2}.
\end{equation}
The corresponding eigenstates~$|{\rm n}\rangle$ are
characterized by four quantum numbers:~$|{\rm n}\rangle=|n,k_x,k_z,s\rangle$,
where~$n$ labels the Landau levels,~$k_x$ and~$k_z$ are wave vector components
and~$s=\pm 1$ label positive and negative energy branches.
The wave functions are~\cite{Witte1987}
\begin{equation} \label{B_Psi}
 \Psi_{\rm n}({\bm r}) \equiv \langle {\bm r} |{\rm n}\rangle =
   \frac{e^{ik_xx+ik_zz}}{4\pi} \phi_n(\xi)
   \left(\begin{array}{c} \mu_{n,k_x,s}^+ \\ \mu_{n,k_x,s}^- \end{array} \right),
\end{equation}
where~$\phi_n(\xi)$ are the harmonic oscillator functions
\begin{equation} \label{B_n}
 \phi_n(\xi) = \frac{1}{\sqrt{L}C_n}{\rm H}_{n}(\xi)e^{-1/2\xi^2},
\end{equation}
in which~${\rm H}_{n}(\xi)$ are the Hermite polynomials and~$C_n=\sqrt{2^n n!\sqrt{\pi}}$.
We defined~$\mu_{n,k_x,s}^{\pm} =\nu_{n,k_x}\pm s/ \nu_{n,k_x}$,
where~$\nu_{n,k_x}=\sqrt{mc^2/E_{n,k_z}}$.

We want to calculate an average packet velocity in a magnetic field.
We can, as before, introduce the Heisenberg picture for the time-dependent
velocity operator. Then the~$j$-th component of the average velocity is,
see Eq.~(\ref{Av_vz}),
\begin{equation} \label{Bv_vj}
\langle \hat{v}_j(t)\rangle = \langle W|\tau_3 e^{i\hH t/\hbar} \hat{v}_j e^{-i\hH t/\hbar}|W\rangle,
\end{equation}
where~$\hat{v}_j =\partial \hH/\partial \hat{p}_j$. For the Hamiltonian~(\ref{B_H})
in the asymmetric gauge we find
\begin{eqnarray}
\hvx &=&(\tau_3 +i\tau_2)\left(\frac{\hat{p}_x - qBy}{m}\right), \label{Bv_vx} \\
\hvy &=&(\tau_3 +i\tau_2)\frac{\hat{p}_y}{m},                    \label{Bv_vy} \\
\hvz &=&(\tau_3 +i\tau_2)\frac{\hat{p}_z}{m}.                    \label{Bv_vz}
\end{eqnarray}
The unity operator is now
\begin{equation} \label{Bv_one}
\hat{1} = \sum_{\rm n}|{\rm n}\rangle\langle{\rm n}|s_n\tau_3,
\end{equation}
where the states~$\langle {\bm r} |{\rm n}\rangle$
are given in Eq.~(\ref{B_Psi}) and~$s_n=\pm 1$
are the quantum numbers associated with the states~$|{\rm n}\rangle$. The proof
of the above identity is given in Appendix~\ref{App_Ident}. Using the unity operator
we expand the packet~$|W\rangle$ in term of the eigenstates of~$\hH$ [see Eq.~(\ref{Av_W})]
\begin{equation} \label{Bv_W}
|W\rangle = \sum_{\rm n}s_{\rm n} |{\rm n}\rangle \langle {\rm n}|\tau_3|W\rangle
\equiv \sum_{\rm n}s_{\rm n} |{\rm n}\rangle W_{\rm n},
\end{equation}
where~$W_{\rm n}= \langle {\rm n}|\tau_3|W\rangle$.
Inserting~$|W\rangle$ into Eq.~(\ref{Bv_vj}) one obtains [see Eq.~(\ref{Av_vz})]
\begin{equation} \label{Bv_vjW1}
\langle \hat{v}_j(t)\rangle =
 \sum_{{\rm n}{\rm m}}s_{\rm n} s_{\rm m} W_{\rm n}^* W_{\rm m}
 \langle {\rm n}|\tau_3e^{i\hH t/\hbar} \hat{v}_j e^{-i\hH t/\hbar}|{\rm m}\rangle.
 \end{equation}
There is~$e^{-i\hH t/\hbar}|{\rm n}\rangle=e^{-i\omega_{\rm n}t}|{\rm n}\rangle$,
where~$\omega_{\rm n} = s_{\rm n}E_{n,k_x}/\hbar$.
Proceeding the same way as in Section~\ref{Sec_Vac} we have
\begin{equation} \label{Bv_tauH}
\langle {\rm n}|\tau_3 e^{i\hH t/\hbar}=\langle {\rm n}|e^{i\hH^{\dagger} t/\hbar}\tau_3
=e^{i\omega_{\rm n}t}\langle {\rm n}|\tau_3,
\end{equation}
which finally gives
\begin{equation} \label{Bv_vjW2}
\langle \hat{v}_j(t)\rangle =
 \sum_{{\rm n}{\rm m}}s_{\rm n} s_{\rm m} W_{\rm n}^* W_{\rm m}
 e^{i(\omega_{\rm n}-\omega_{\rm m})t} \langle {\rm n}|\tau_3\hat{v}_j|{\rm m}\rangle.
 \end{equation}
The matrix elements of velocity operators calculated between
the states~$|{\rm n}\rangle$,~$|{\rm m}\rangle$ are
\begin{eqnarray}
 \langle {\rm n}|\tau_3\hvy|{\rm m}\rangle &=& c\frac{\lambda_c}{i\sqrt{2}L}
 \nu_{n,k_z}\nu_{m,k_z}\delta_{k_x,k_x'}\delta_{k_z,k_z'}
 \times \nonumber \\
                && (\sqrt{n+1}\delta_{m,n+1} - \sqrt{n}\delta_{m,n-1}), \\
 \langle {\rm n}|\tau_3\hvx|{\rm m}\rangle &=& c\frac{\lambda_c}{\sqrt{2}L}
    \nu_{n,k_z}\nu_{m,k_z}\delta_{k_x,k_x'}\delta_{k_z,k_z'}
  \times \nonumber \\
                && (\sqrt{n+1}\delta_{m,n+1} + \sqrt{n}\delta_{m,n-1}), \\
 \langle {\rm n}|\tau_3\hvz|{\rm m}\rangle &=& \frac{p_z}{m}\nu_{n,k_z}\nu_{m,k_z}
     \delta_{k_x,k_x'}\delta_{k_z,k_z'}\delta_{m,n}.
\end{eqnarray}
The matrix elements of~$\hvy$ and~$\hvx$ are nonzero for the states with~$m=n\pm 1$
and arbitrary indexes~$s_n$ and~$s_m$. The matrix elements of~$\hvz$ are nonzero for~$m=n$
and arbitrary indexes~$s_n$ and~$s_m$.
To simplify the further analysis we assume the initial wave packet~$W({\bm r})$
to be in a separable form
\begin{equation}
W({\bm r}) = W_{xy}(x,y)W_z(z).
\end{equation}
Then there is
\begin{equation}
W_{\rm n}=\langle {\rm n}|\tau_3| W\rangle = \mu_{n,k_z}^+ g_z(k_z)F_{n}(k_z),
\end{equation}
where
\begin{equation}
 F_n(k_x) = \frac{1}{\sqrt{2L}C_n} \int_{-\infty}^{\infty} g_{xy}(k_x,y)e^{-\frac{1}{2}\xi^2}{\rm H}_{n}(\xi)dy,
\end{equation}
in which
\begin{equation}
 g_{xy}(k_x,y) = \frac{1}{\sqrt{2\pi}} \int_{-\infty}^{\infty} w_{xy}(x,y)e^{ik_xx} dx,
\end{equation}
and
\begin{equation}
 g_z(k_z) = \frac{1}{\sqrt{2\pi}} \int_{-\infty}^{\infty} w_{z}(z)e^{ik_zz} dz.
\end{equation}
For~$\langle \hat{v}_y(t)\rangle$ we obtain
\begin{eqnarray*}
\langle \hat{v}_y(t)\rangle = c\frac{\lambda_c}{2\sqrt{2}iL}\sum_{n,m=0}^{\infty} \int_{-\infty}^{\infty} dk_z \times
       \nonumber \\
  |g_z(k_z)|^2 \left(\sqrt{n+1}\delta_{m,n+1} -\sqrt{n}\delta_{m,n-1} \right)U_{n,m} \times
       \nonumber \\
       \left\{(1+\nu_m^2\nu_n^2)\cos(\omega_{m}t-\omega_{n}t) + (\nu_m^2\nu_n^2-1)\cos(\omega_{m}t+\omega_{n}t)
       \right. \nonumber \\ \left.
            + i(\nu_m^2+\nu_n^2)\sin(\omega_{m}t-\omega_{n}t)+i(\nu_m^2-\nu_n^2)\sin(\omega_{m}t+\omega_{n}t)
 \right\}.
\end{eqnarray*}
In the above expressions we use the notation~$\nu_{n}\equiv \nu_{n,kz}$
and~$\omega_{n} = E_{n,k_z}/\hbar$, and
\begin{equation}
 U_{n,m} = \int_{-\infty}^{\infty} F^*_n(k_x)F_{m}(k_x)d k_x.
\end{equation}
For a Gaussian packet of Eq.~(\ref{P_wr}) one can obtain analytical
expressions for~$U_{n,m}$, see Appendix~\ref{App_Gauss}.
After performing the summation over~$m$ and changing~$n\rightarrow n+1$ in~$\delta_{m,n-1}$ terms
we finally obtain
\begin{eqnarray} \label{Bv_vyt}
\langle \hat{v}_y(t)\rangle = -c\frac{\lambda_c}{2\sqrt{2}L}\sum_{n=0}^{\infty}\sqrt{n+1}(U_{n+1,n}+U_{n,n+1})
\nonumber \\
 \int_{-\infty}^{\infty} |g_z(k_z)|^2 \left\{(\nu_{n+1}^2+\nu_n^2)\sin(\omega_{n+1}t-\omega_{n}t) \right. \nonumber \\ \left.
        +(\nu_{n+1}^2-\nu_n^2)\sin(\omega_{n+1}t+\omega_{n}t) \right\} dk_z,
\end{eqnarray}
\begin{eqnarray} \label{Bv_vxt}
\langle \hat{v}_x(t)\rangle = -c\frac{\lambda_c}{2\sqrt{2}L}\sum_{n=0}^{\infty}\sqrt{n+1}(U_{n+1,n}+U_{n,n+1})
\nonumber \\
 \int_{-\infty}^{\infty} |g_z(k_z)|^2 \left\{(1+\nu_{n+1}^2\nu_n^2)\cos(\omega_{n+1}t-\omega_{n}t) \right. \nonumber \\ \left.
        + (1-\nu_{n+1}^2\nu_n^2)\cos(\omega_{n+1}t+\omega_{n}t) \right\} dk_z,
\end{eqnarray}
\begin{eqnarray} \label{Bv_vzt}
\langle \hat{v}_z(t)\rangle = \frac{c\lambda_c}{2}\sum_{n=0}^{\infty}U_{n,n} \int_{-\infty}^{\infty}
 k_z |g_z(k_z)|^2 \times
\nonumber \\
 \left\{(1+\nu_{n}^4) + (1-\nu_{n}^4)\cos(2\omega_{n}t) \right\} dk_z.
\end{eqnarray}

Equations~(\ref{Bv_vyt}) -~(\ref{Bv_vzt}) are our final results for the average velocity of
wave packet in a magnetic field. Both the arguments of sine and cosine functions
as well as coefficients~$\nu_{n}$ and~$\nu_{n+1}$
depend on~$k_z$, so all the integrals vanish in the limit~$t\rightarrow \infty$
as a consequence of the Riemann-Lebegues theorem and the resulting oscillations
have a transient character. The velocity of the packet oscillates with many
frequencies~$\omega_{n+1}\pm\omega_{n}$ (or~$2\omega_{n}$ for~$\hat{v}_z$),
but in practice the spectrum is limited to a few frequencies related to the largest
coefficients~$U_{n+1,n}$ and~$U_{n,n}$. The frequencies~$\omega_{n+1}-\omega_{n}$ correspond to
the intraband transitions and they can be interpreted as the cyclotron resonances.
These frequencies do not appear in~$\hat{v}_z$ velocity. On the other hand,
the frequencies~$\omega_{n+1}+\omega_{n}$ and~$2\omega_{n}$ (for~$\hat{v}_z$)
correspond to interband transitions and they can be interpreted as the
Zittebewegung components of the motion in analogy to the situation at zero field.
The motion in the~$x-y$ directions requires that~$k_{0x}\neq 0$ because
for~$k_{0x}=0$ all the coefficients~$U_{n+1,n}$ and~$U_{n,n+1}$ vanish~\cite{Rusin2010}.
For the motion in the~$z$ direction one needs only that~$k_{0z}\neq 0$,
because the coefficients~$U_{n,n}$ are nonzero for any~$k_{0x}$ vector~\cite{Rusin2010}.

Considering the non-relativistic limit in Eqs.~(\ref{Bv_vyt}) -~(\ref{Bv_vzt}),
there is~$\hbar\omega_c \ll mc^2$ and~$\hbar k_z \ll mc$,
so that~$\omega_{n+1}-\omega_{n} \simeq \hbar\omega_c$
and~$\omega_{n+1}+\omega_{n} \simeq 2mc^2/\hbar$. In this limit
there is~$\nu_{n+1}\simeq \nu_{n} \simeq 1$,
and the ZB part of velocity is nearly zero. In this case we may decouple
in Eqs.~(\ref{Bv_vyt}) -~(\ref{Bv_vzt}) the summation over~$n$ and
integration over~$k_z$. This gives~\cite{Rusin2010}
\begin{eqnarray}
 \sum_{n=0}^{\infty}\sqrt{n+1}U_{n+1,n} &=& -\frac{k_{0x}L}{\sqrt{2}} \, \\
 \sum_{n=0}^{\infty} U_{n,n} &=& 1. \label{BV_sum_Unn}
\end{eqnarray}
Integrating over~$k_z$ one gets
\begin{eqnarray}
 \label{Bv_vy_small} \langle \hat{v}_y(t)\rangle &\simeq & \frac{\hbar k_{0x}}{m}\sin(\omega_ct),\\
 \label{Bv_vx_small} \langle \hat{v}_x(t)\rangle &\simeq & \frac{\hbar k_{0x}}{m}\cos(\omega_ct),\\
 \label{Bv_vz_small} \langle \hat{v}_z(t)\rangle &\simeq & \frac{\hbar k_{0z}}{m}.
\end{eqnarray}
\begin{figure}
\includegraphics[width=8cm,height=8cm]{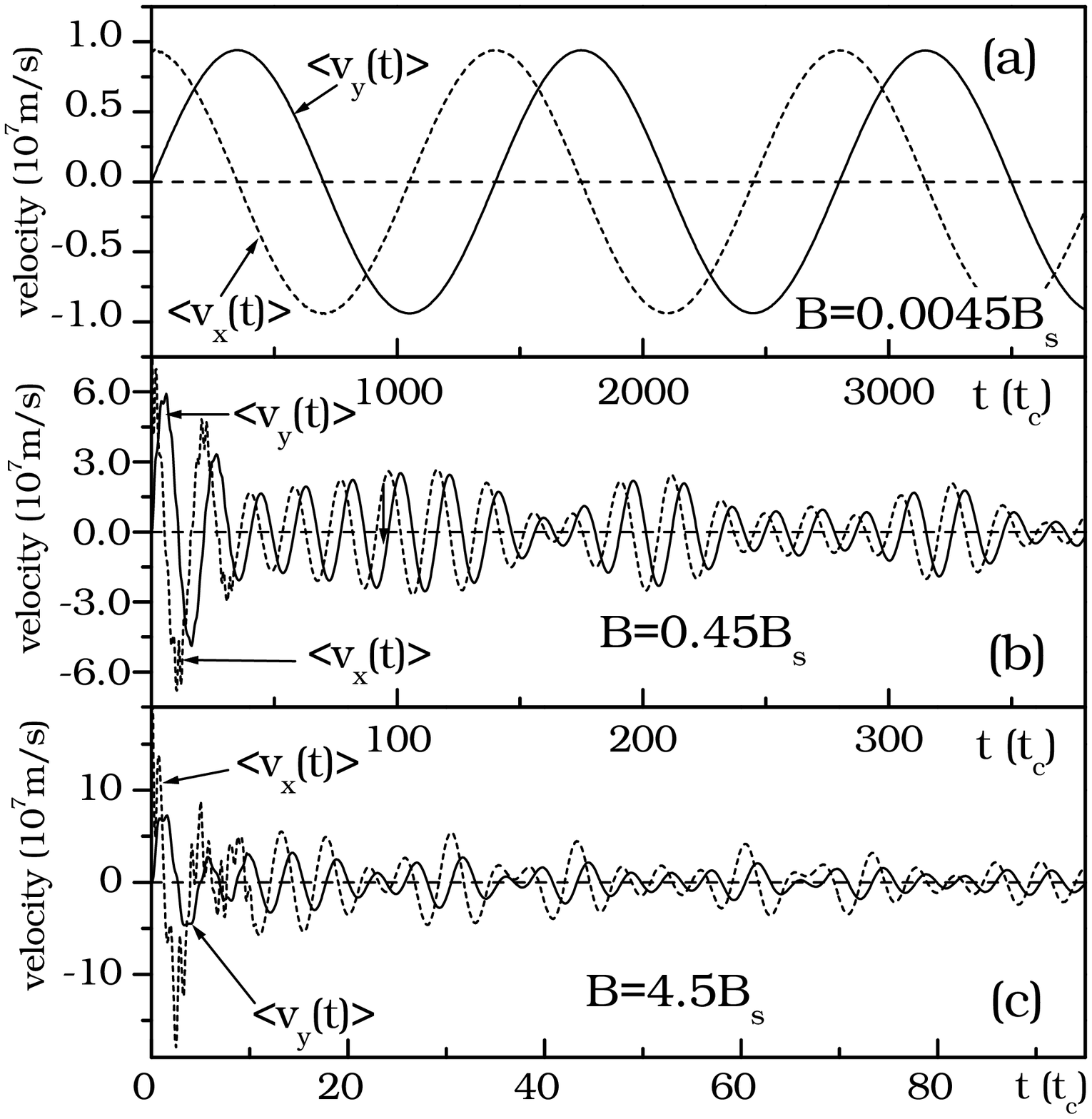}
\caption{\label{Figure3} Time dependent velocity components for ellipsoidal wave packet
         at various magnetic fields.
         For low fields, see~(a), cyclotron motion is obtained, for higher fields,
         see~(b) and~(c), packet velocity includes both cyclotron and Zitterbewegung
         frequencies. In all cases the motion decays in time.}
\end{figure}

\begin{figure}
\includegraphics[width=8cm,height=8cm]{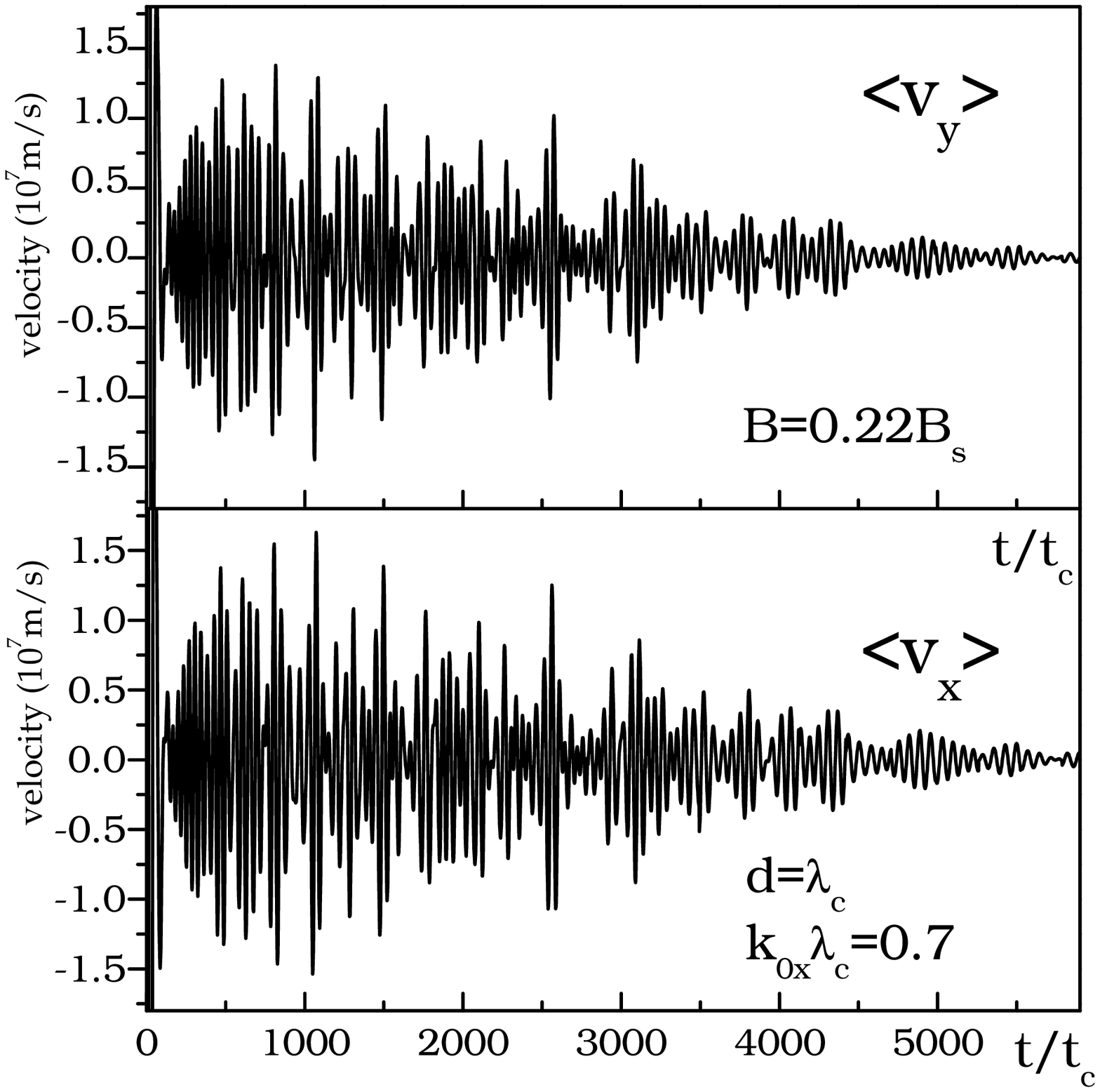}
\caption{\label{Figure4} Average velocity of the spherical wave packet
         in longer time scale. The collapse-revival patterns are seen in ZB oscillations.
         The motion has a transient character.}
\end{figure}

\begin{figure}
\includegraphics[width=8cm,height=8cm]{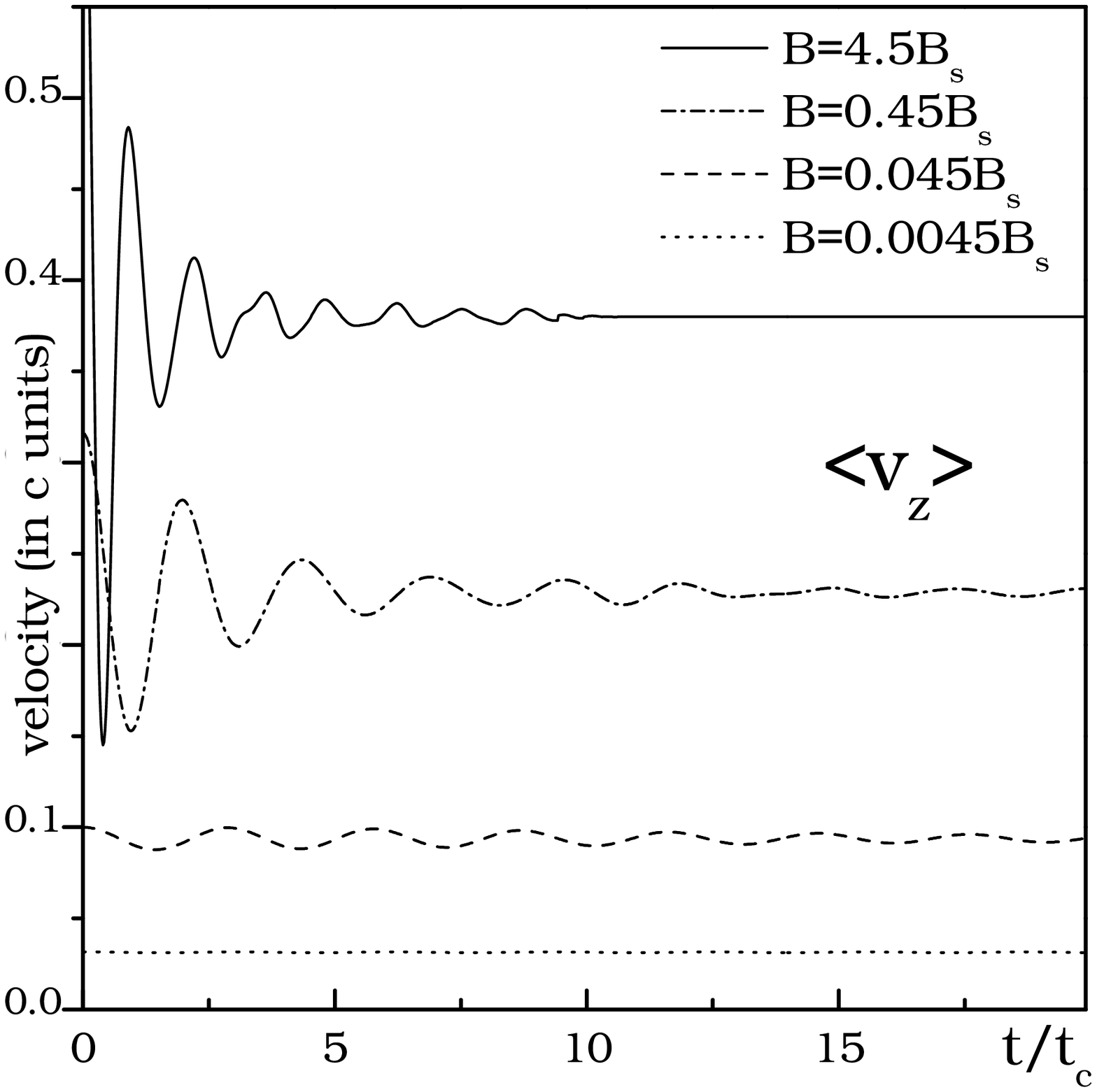}
\caption{\label{Figure5} Average packet velocity~$\langle \hat{v}_z(t)\rangle$
         in the direction parallel to magnetic field versus time for four
         values of~$B$. Transition to the non-relativistic limit is visible.
         Parameters are the same as those used for Fig.~\ref{Figure3}.}
\end{figure}
Thus in the non-relativistic limit the particle moves on a circular orbit with the cyclotron
frequency in the~$x-y$ plane and a constant velocity in the~$z$ direction.
Let us introduce a measure of intensity of a magnetic field by its relation to an
effective Schwinger field~$\hbar eB_s/m = mc^2$ or, equivalently, by~$L_s=\hbar/mc$.
There is~$B_s=4.41 \times 10^9(m/m_e)^2$~T, where~$m_e$ is the electron mass.
Below we perform calculations for pions~$\pi^+$ having the
mass~$m\simeq 273.1$~$m_e$,
so the effective Schwinger field is~$B_s=3.29\times 10^{14}$~T.

In Fig.~\ref{Figure3} we plot the average packet velocity for three values of magnetic field.
The ellipsoidal packet is selected with a nonzero initial momentum~$k_{0x}$.
We assume that the five parameters:~$d_x$,~$d_y$,~$d_z$,~$L$ and~$k_{0x}^{-1}$
have similar orders of the magnitude which are the optimal conditions for the
appearance of Zitterbewegung phenomenon. In Fig.~\ref{Figure3} we selected
parameters:~$d_x=0.91(B_s/B)\lambda_c$,~$d_y=0.82(B_s/B)\lambda_c$,~$d_z=0.68(B_s/B)\lambda_c$,
$k_{0x}=0.7(B/B_s)\lambda_c^{-1}$ and~$k_{0z}=0$, where~$B_s$ is the effective Schwinger field.
For~$B=4.5B_s$ we set~$k_{0x}=\lambda_c^{-1}$.
For low fields ($B=0.0045 B_s$) the packet moves on a circular orbit,
see Eqs.~(\ref{Bv_vy_small}) and~(\ref{Bv_vx_small}). For such fields the ZB components of the motion
are negligible. For higher fields the packet motion includes both the intraband
and interband~(ZB) components so that several frequencies give significant
contributions to the motion. In all cases the motion has a transient character
but for low fields its decay time is very long.
In Fig.~\ref{Figure4} we plot components of average velocity of a spherical packet in a longer
time scale. The collapse-and-revival patterns occur for both velocity components.
After a sufficiently long time the oscillations disappear.
In Fig.~\ref{Figure5} we show the average velocity~$\langle \hat{v}_z(t)\rangle$
of an ellipsoidal packet having the same parameters as those used in Fig.~\ref{Figure3}.
For large magnetic fields the
motion in the~$z$ direction is similar to that in the field-free case
exhibiting ZB oscillations, see Fig.~\ref{Figure1}. For smaller fields the ZB
oscillations disappear and only the classical motion remains, see Eq.~(\ref{Bv_vz_small}).
Finally, it should be mentioned that in the two-dimensional case the ZB oscillations
do not disappear in time~\cite{Rusin2010}.

\section{Simulation of ZB \label{Sec_Sim}}

The phenomenon of Zitterbewegung for relativistic particles in a vacuum has an unfavorable
high frequency corresponding to the energy gap between the positive and negative energy
branches:~$\hbar\omega_0 \simeq 2mc^2$, and a very small amplitude on the order of
the effective Compton wavelength~$\Delta {\bm r} \simeq \hbar/(mc)$, see Eq.~(\ref{H_rz_11q}).
Thus, similarly to the case of relativistic electrons, one can not hope at present
to observe directly the ZB in a vacuum. However, it was recently demonstrated by
Gerritsma {\it et al.} that one can simulate the ZB of electrons in a vacuum using trapped
ions interacting with laser beams~\cite{Gerritsma2010}. In this experiment the authors
simulated the linear momentum~$\hat{p}_i$ appearing in the Dirac equation with the
use of Jaynes-Cumminngs interaction between the electrons on trapped ion levels
and the electromagnetic radiation. The decisive advantage of such a simulation is that one
can tailor the frequency and amplitude of ZB making them considerably more favorable than
the values for a vacuum. Clearly, it would be of interest to simulate the ZB of
a Klein-Gordon particle using similar methods. The problem is that in KGE one deals with
{\it squares} of momentum components~$\hat{p}^2$, which are more difficult to simulate
with the Jaynes-Cumminngs interaction. For this reason we choose a different route.

The Klein-Gordon equation appears in several {\it classical} systems, usually as a modification
of the wave equation~$\Box\phi=0$. Under some conditions KGE is used to describe
sound waves in ducts~\cite{Forbes2003,Forbes2004},
electromagnetic waves in the ionosphere~\cite{JacksonBook,Tsynkov2009},
transverse modes of wave guides~\cite{GeyiBook} and oceanic waves~\cite{Wurmser1997}.
Below we examine in more detail a model proposed by Morse and Feshbach
in which one can simulate KGE with the use of a piano string and a thin rubber
sheet~\cite{MorseBook}. Employing this example we demonstrate similarities and differences
between ZB in the relativistic KGE and its classical analogues.

\begin{figure}
\includegraphics[width=8cm,height=8cm]{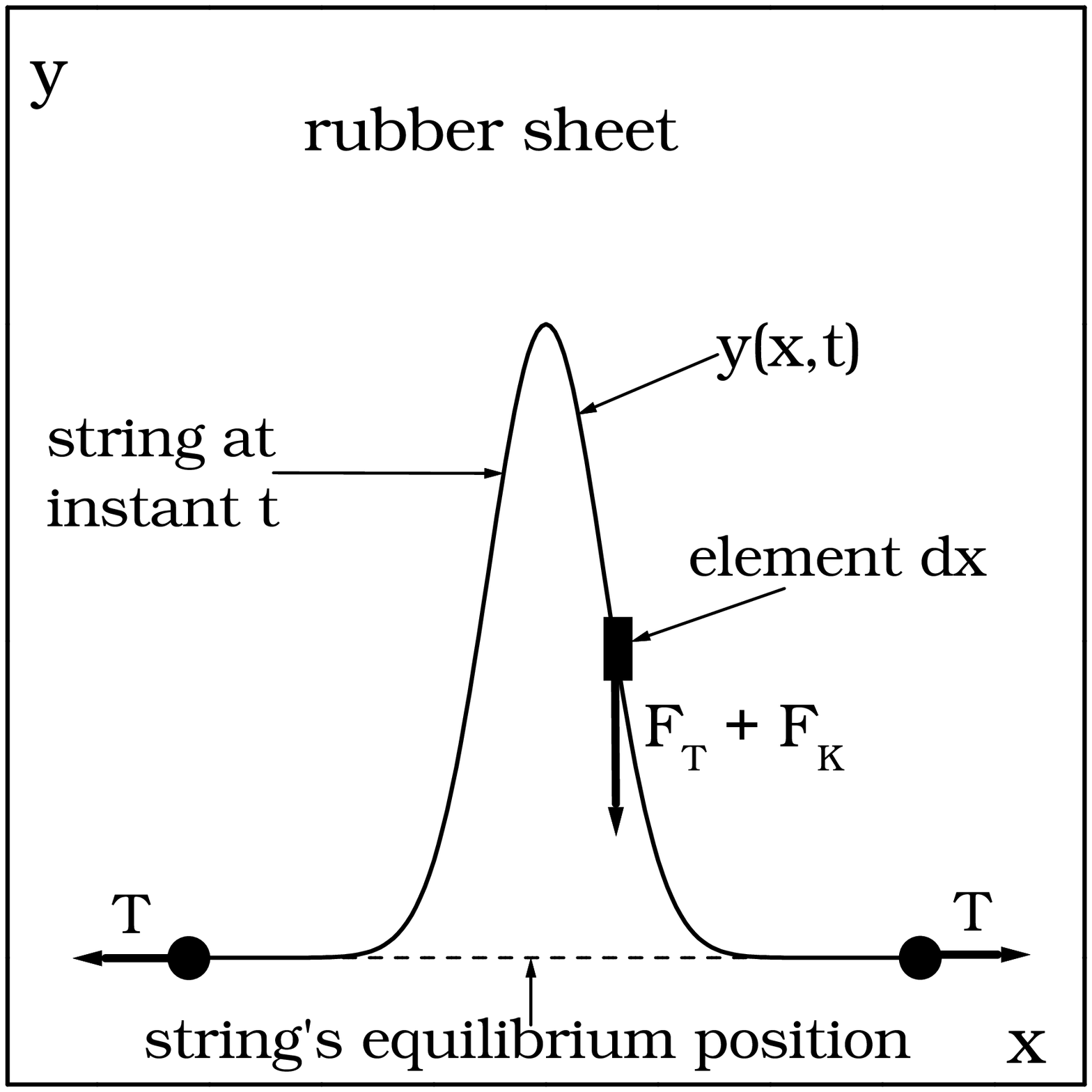}
\caption{\label{Figure6}
         Classical simulation of KGE according to Morse and Feshbach~\cite{MorseBook}.
         Flexible string is anchored at two points and tension~$T$ is applied to each end.
         The string is also attached to a thin rubber sheet.
         At instant~$t$ the shape of string is given by~$y(x,t)$.
         There are two forces acting on each element~$dx$ of the string:
         restoring force~$F_T$ due to applied tension
         and elastic force~$F_K$ of stretched rubber.}
\end{figure}

Let us consider flexible one dimensional string in the~$x$ direction, see Fig.~\ref{Figure6}.
We assume that the string is uniform with a linear density~$\rho$.
A uniform tension~$T$ is applied to each element~$dx$ of the string.
We neglect all other forces acting on the string (e.g. gravity) and the stiffness of the string.
Let~$y(x,t)$ be a displacement of the element~$dx$ of the string from its equilibrium
position at an instant~$t$. We assume that~$y(x,t)$ is small compared to the length of the string
and to the distances to each end of the string. The restoring force acting on
each element~$dx$ of the string is~$F_T=Tdx(\partial^2 y/\partial x^2)$ and
displacement~$y(x,t)$ of the released string changes according to
the wave equation~\cite{MorseBook}
\begin{equation}
\frac{1}{u^2} \frac{\partial^2 y}{\partial t^2} = \frac{\partial^2 y}{\partial x^2},
\end{equation}
where~$u^2=T/\rho$. Now we attache the string to an elastic substrate, e.g.
to a thin sheet of rubber which can shrink or expand in the~$y$ direction.
Then, in addition to the restoring force due to the tension, there will be another restoring
force due to the elastic rubber acting on each element of the string.
If the element~$dx$ is displaced to~$y(x,t)$ and the rubber sheet obeys the Hook law,
the restoring force acting on the element~$dx$ of the string
is~$F_K(x,t)=-Ky(x,t)dx$, where~$K$ is the elastic
constant of the rubber sheet. The second Newton law for the element~$dx$ of the string
having mass~$dm=dx\rho$ is~$dx\rho (\partial^2 y/\partial t^2) = F_T + F_K$,
so the equation of motion of the released string is
\begin{equation} \label{S_KG}
 \frac{1}{u^2} \frac{\partial^2 y}{\partial t^2} = \frac{\partial^2 y}{\partial x^2} - \nu^2y,
\end{equation}
where~$\nu^2=K/T$. Equation~(\ref{S_KG}) has the form of wave KGE
with the light speed replaced by~$u$ and the mass term~$m^2c^2/\hbar^2$ replaced by~$\nu^2$.
Comparing Eq.~(\ref{S_KG}) with Eq.~(\ref{WE_KG})
we find the following correspondence between parameters of the two systems
\begin{eqnarray} \label{S_an_c2}
\frac{T}{\rho} &\leftrightarrow& c^2, \\
\frac{K}{T}    &\leftrightarrow& \frac{m^2c^2}{\hbar^2} = \lambda_c^{-2}. \label{S_an_lc2}
\end{eqnarray}
Thus one can simulate values of~$c$ and~$\lambda_c$ by changing material
parameters~$\rho$,~$K$ and~$T$.

However, there exist also limitations of such a simulation
and they affect a possibility of observation of ZB motion in classical analogues of KGE.
The first difference between the relativistic KGE and its classical counterpart is that
the wave function~$\phi$ in the relativistic KGE is not an observable. On the other
hand, all classical analogues of~$\phi$ (such as a displacement of the string,
the pressure of sound or the oceanic waves, the intensity of electromagnetic field etc.)
are observable quantities. The second
difference is that the relativistic function~$\phi$ is a function of complex variable,
while its classical counterpart is a function of real variable.
A direct consequence of these limitations for observation of ZB
in classical systems is that, for any real function~$\xi(\bm r,t)$ being the solution of KGE,
the current density associated with this function is always
zero:~$ {\bm j} \propto \left[\xi^*\nabla \xi - (\nabla \xi^*)\xi \right] =0$.
Therefore we are not able to simulate directly the current or velocity oscillations
calculated in the previous sections.

\begin{figure}
\includegraphics[width=8cm,height=8cm]{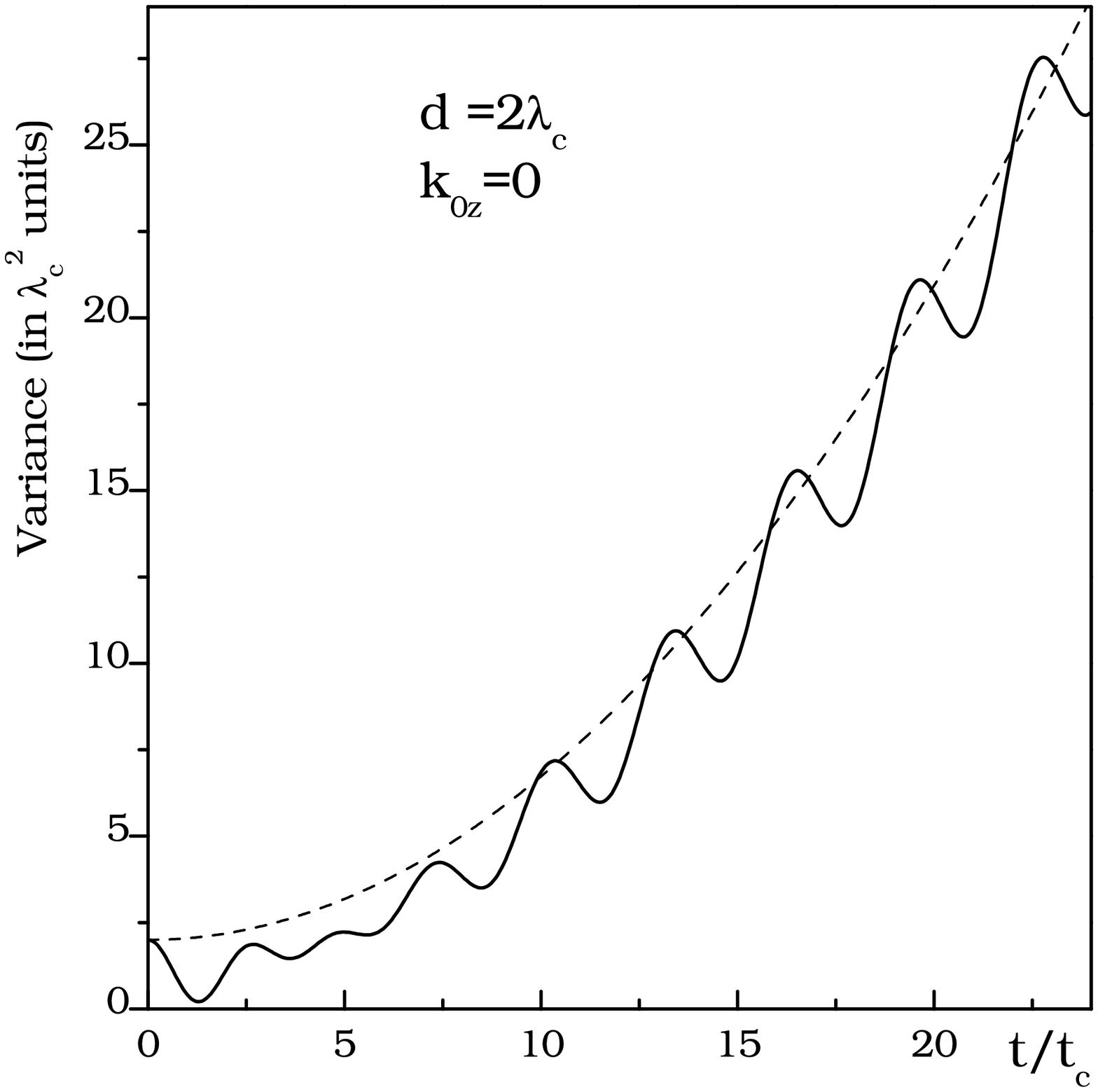}
\caption{\label{Figure7} Calculated classical variance of position of wave packet
         propagating according to KGE.
         Dashed line: non-oscillating part of variance;
         solid line: total variance. For string oscillations analyzed
         in the text there is~$\lambda_c^s=4.47$~mm and~$t_c^s=2.37\times 10^{-5}$~s.}
\end{figure}

To overcome this problem let us consider the motion of a {\it neutral} particle
described by a real field~$\xi$. For simplicity we assume a one-dimensional KGE that can be
simulated by a flexible string attached to an elastic substrate described above.
In our calculations we use the relativistic form of KGE but the final results will
be presented for parameters corresponding to the flexible string model.
We assume the initial wave packet to be a real Gaussian function without an initial momentum
\begin{equation} \label{S_wr}
 w_0(x)=\frac{1}{(d\sqrt{\pi})^{1/2}} \exp[-x^2/(2d^2)].
\end{equation}
Its Fourier transform is
\begin{equation} \label{S_wk}
 w_0(k)=(2d\sqrt{\pi})^{1/2}\exp[-d^2k^2/2].
\end{equation}
A real solution~$\xi(x,t)$ of KGE is
\begin{equation} \label{S_xi}
 \xi(x,t) = \frac{1}{2\pi}\int_{-\infty}^{\infty}w_0(k)\cos(kx-\omega_kt) dk,
\end{equation}
where~$\omega_k=\omega_0\sqrt{1+(k\lambda_c)^2}$. The average current
for the real wave packet of Eq.~(\ref{S_wr}) is zero and no ZB occurs.
Thus we turn to other physical operators which do not commute with
the KG Hamiltonian~(\ref{H_KG}). Namely, we calculate a {\it variance}
of the position operator for the above real function~$\xi(x,t)$
\begin{equation} \label{S_V_def}
 V = \langle \xi|\hat{x}^2|\xi\rangle - \langle \phi|\hat{x}|\phi\rangle^2 =
     \langle \xi|\hat{x}^2|\xi\rangle,
\end{equation}
since~$\langle \phi|\hat{x}|\phi\rangle=0$.
Assuming~$\xi(x,t)$ in the form~(\ref{S_xi}) we have
\begin{eqnarray} \label{S_V}
V  &=& \iiint_{\infty}^{\infty} w_0(k)w_0(k')\cos(kx-\omega_kt) \times \nonumber \\
   & & \cos(k'x-\omega_{k'}t)x^2 dx dk dk' = \nonumber \\
   & & \iiint_{\infty}^{\infty}
       \left[B_kB_{k'}e^{ix(k+k)}+ B_kB_{k'}^*e^{ix(k-k')} \right. \nonumber \\
   & &\left. B_k^*B_{k'} e^{-ix(k-k')} + B_k^*B_{k'}^*e^{-ix(k+k')} \right]x^2 dx dk dk', \ \ \ \ \
\end{eqnarray}
where~$B_k=w_0(k)e^{-i\omega_kt}/(4\pi)$.
Consider the first of the four terms given above.
Because~$w_0(k)$ and~$B_k$ decay exponentially for~$k \rightarrow \pm\infty$,
one can change the order of integration over~$x$,~$k$ and~$k'$ and
replace~$x^2\rightarrow (\partial/\partial ik)(\partial/\partial ik')$. Then we
integrate by parts over~$k$ and~$k'$ and obtain
\begin{eqnarray} \label{S_BB}
\iiint_{\infty}^{\infty} B_kB_{k'}e^{ix(k+k)}x^2 dx dk dk' \nonumber \\
= \iiint_{\infty}^{\infty} \frac{\partial B_k}{\partial k}
        \frac{\partial B_{k'}}{\partial k'}e^{ix(k+k)} dx dk dk' \nonumber \\
 = 2\pi\int_{\infty}^{\infty}\left. \frac{\partial B_k}{\partial k}
                             \frac{\partial B_{k'}}{\partial k'}\right|_{k'=k} dk.
\end{eqnarray}
The other three terms in Eq.~(\ref{S_V}) are calculated similarly.
After some manipulations we find
\begin{equation} \label{S_V_wyn}
 V = V_1^c + V_1^{osc} + V_2^c + V_2^{osc} + V_3,
\end{equation}
where
\begin{eqnarray}
\label{S_V1c}
V_1^c     &=& \frac{d^3}{2\sqrt{\pi}}\int_{\infty}^{\infty} e^{-d^2k^2} (kd)^2 dk, \\
\label{S_V1osc}
V_1^{osc} &=& \frac{d^3}{2\sqrt{\pi}}\int_{\infty}^{\infty} e^{-d^2k^2} (kd)^2 \cos(2\omega_kt) dk,\\
\label{S_V2c}
V_2^c     &=& \frac{d(ct)^2}{2\sqrt{\pi}}\int_{\infty}^{\infty}\frac{ e^{-d^2k^2}(k\lambda_c)^2}{1+(k\lambda_c)^2} dk, \\
\label{S_V2osc}
V_2^{osc} &=& -\frac{d(ct)^2}{2\sqrt{\pi}}\int_{\infty}^{\infty} \frac{ e^{-d^2k^2}(k\lambda_c)^2}{1+(k\lambda_c)^2} \cos(2\omega_kt)dk, \\
\label{S_V3}
V_3       &=& \frac{d^2(ct)}{\sqrt{\pi}}\int_{\infty}^{\infty} \frac{ e^{-d^2k^2}(k\lambda_c)}{\sqrt{1+(k\lambda_c)^2}}\sin(2\omega_kt) dk.
\end{eqnarray}
The term~$V_3$ is odd in~$k$ and it vanishes upon the integration.
For~$t=0$ the variance in Eq.~(\ref{S_V_wyn}) is equal to the variance~$V_0=d^2/2$
of the initial packet~$w_0(x)$.
The variance given in Eq.~(\ref{S_V_wyn}) consists of oscillating and non-oscillating
terms. For large times the non-oscillating terms grow in time as~$d^2/2 + Ct^2$,
where~$C$ is a constant depending on~$d$. The quadratic dependence of the variance
on time is similar to that of a Gaussian wave packet in the non-relativistic quantum mechanics.
The oscillations in Eqs.~(\ref{S_V1osc}) and~(\ref{S_V2osc}) have the same interband
frequency~$2\omega_k$ as the velocity oscillation given in Eq.~(\ref{H_vz_11q}).
Therefore the oscillations of variance of the position operator can be interpreted as a signature
of Zitterbewegung in classical systems. The term~$V_1^{osc}$
has a decaying character and it vanishes after a few oscillations.
The~$V_2^{osc}$ term gives persistent oscillations
because of the presence of~$t^2$ factor in front of the integral.
To estimate the time dependence of these oscillations
we consider the limit of large packet widths~$d\gg \lambda_c$. In this case the Gaussian
function in Eq.~(\ref{S_V2osc}) restricts the integration
to small values of~$k$. Then we may disregard~$(k\lambda_c)^2$ term in the denominator of
integrand and expand~$\omega_k$ under the cosine function.
This gives approximately
\begin{eqnarray} \label{S_V_Appr0}
V_2^{osc} &\simeq&
-\frac{d(ct)^2}{2\sqrt{\pi}}\int_{\infty}^{\infty} e^{-d^2k^2}(k\lambda_c)^2
   \cos[\omega_0t(2+k^2\lambda_c^2)] dk \nonumber \\
  &=&-\frac{d(ct)^2}{4}\sum_{\eta=\pm 1} \frac{e^{2i\eta\omega_0t}}{(d^2+i\eta\omega_0t)^{3/2}}.
\end{eqnarray}
For large time we may approximate in Eq.~(\ref{S_V_Appr0})
\begin{equation} \label{S_V_Appr}
 V_2^{osc}\simeq -C_dt^{1/2}\cos(2\omega_0t),
\end{equation}
where~$C_d$ is a constant depending on~$d$. Thus the oscillations
of variance are persistent, their amplitude increases with time as~$t^{1/2}$ and their
frequency is~$2\omega_0$. Since non-oscillating terms~$V_2^c$ increase as~$t^2$, the
total variance of the packet has a quadratic time dependence with
superimposed oscillations. This behavior is illustrated in Fig.~\ref{Figure7}.
In our classical considerations we do not face the problem of negative variances
that can occur for some quantum systems, see Refs.~\cite{Lev2002,Semenov2008}.
For~$t<5t_c$ the oscillations
have an irregular character because of the contribution of~$V_1^{osc}$ term.

Estimating the characteristic frequency~$2\omega_0$ for the flexible string attached to elastic
substrate we have
\begin{equation} \label{S_an_omega0}
 \omega_0^2 = \frac{m^2c^4}{\hbar^2} = c^2 \times \frac{1}{\lambda_c^2}
  \longleftrightarrow \left(\frac{T}{\rho}\right) \left(\frac{K}{T} \right) = \frac{K}{\rho},
\end{equation}
so that the analogue of the relativistic frequency~$\omega_0$ does not
depend on the applied tension. Taking a piano copper string of the bulk
density~$\rho_{3D}=8940$~kg/m$^3$ and having cross section of radius~$r=1$~mm
one gets a linear density~$\rho=\pi r^2\rho_{3D}=2.81\times 10^{-2}$~kg/m.
We identify the rubber elastic constant~$K$ with the Young
modulus~$K=0.05\times 10^9$ N/m$^2$.
Then the analogue of ZB frequency given in Eq.~(\ref{S_an_omega0})
is~$2\omega_0^s=8.44\times 10^4$~s$^{-1}$, i.e.
the corresponding frequency is~$f_0=13.43$~kHz, which can be heard by the human ear.
The characteristic time of ZB oscillations is~$t_c^s=1/\omega_0^s=2.37\times 10^{-5}$~s.
Assuming the tension of the string~$T=1000$~N we find from Eq.~(\ref{S_an_lc2}) that
the simulated Compton wavelength is~$\lambda_c^s=4.47$~mm. The initial wave packet should have
widths~$d$ on the order of a few~$\lambda_c^s$, i.e. of a few centimeters,
and it will move with the velocity~$u=188.7$~m/s, see Eq.~(\ref{S_an_c2}).
Thus, it is really possible to simulate and observe the Zitterbewegung
phenomenon in this system. Finally, we observe that in classical simulations
all the involved quantities are well defined observables. Since classical KGE
does not reproduce but only {\it simulates} the quantum KGE, we are allowed to consider
quantities which are not well defined in the quantum world.

\section{Discussion \label{Sec_Disc}}

Our main results for ZB of KG particles in absence of fields are shown in Fig.~\ref{Figure1}
and in the presence of a magnetic field in Figs.~\ref{Figure3} -~\ref{Figure5}.
It is not our purpose here to
consider difficulties of the one-particle Klein-Gordon equation but we keep them in mind.
In particular, we do not consider particle trajectories as they are believed to be not well
defined, see~\cite{SchweberBook}. On the other hand, we describe average particle velocities
and currents both in the Hamiltonian and wave formalisms. The results can be compared to those
for relativistic electrons in a vacuum described by the Dirac equation as well as for electrons
in solids.

Similarly to the Dirac electrons, the ZB phenomenon of KG particles is due to the interference of
positive and negative energy states. In the non-relativistic limit one of the two components
progressively vanishes and the ZB contribution to the motion disappears. This can be clearly
seen in Figs.~\ref{Figure3} and~\ref{Figure5} as well as in Fig.~3 of Ref.~\cite{Rusin2010} for
the Dirac electrons. If particles are described by wave packets the ZB motion decays in time,
see our Fig.~\ref{Figure1} for KE particles and Fig.~2 of Ref.~\cite{Merkl2008} for the Dirac
electrons. This is a general consequence of the Riemann-Lesbegues theorem, as indicated
by Lock~\cite{Lock1979}, calculated by the present authors~\cite{Rusin07b} and experimentally
confirmed by Gerritsma {\it et al.}~\cite{Gerritsma2010}. In all cases the basic frequency of
ZB oscillations is given by the energy difference between the positive and
negative energy branches:~$\hbar\omega_Z \simeq 2mc^2$ with the corresponding particle mass.
The main difference with the Dirac electrons is the spin. For KG particles the interband ZB
frequencies in a magnetic field do not include the spin energies, one does not deal with
the Fermi sea for the negative energy branches, etc. The KG Hamiltonian is quadratic in momenta
which does not allow a direct simulation with the use of Jaynes-Cumminngs interaction.

As to the Zitterbewegung of electrons in narrow-gap semiconductors
and in particular in zero-gap monolayer graphene,
one should emphasize that, although it is also described using a two-band model of
band-structure~\cite{Zawadzki2011}, its physical nature is completely different from ZB
of particles in a vacuum. The ZB in semiconductors or in graphene results from the electron
motion in a periodic potential~\cite{Zawadzki2010}. In zero-gap situation in graphene the
ZB frequency is given by the difference of energies between positive
and negative energy bands corresponding to the average value of
quasi-momentum~$\hbar {\bm k}_0$ for the wave packet~\cite{Rusin07b}.
A one-dimensional system which strongly resembles the KG particle in a vacuum is presented by
electrons in carbon nanotubes: one can neglect the electron spin dealing with an energy gap
controlled by the tube's diameter~\cite{Zawadzki2006}.
The resulting ZB frequency and amplitude have values easily accessible experimentally.
On the other hand, it is at present not clear how to
follow dynamics of a single electron in a solid.
As to KG particles in a vacuum, one is bound to recourse to simulations
since the ZB frequency and amplitude as well as field intensities
necessary to see ZB effects in the presence of a magnetic field,
exceed the present experimental possibilities.

We present a classical simulation of ZB by using a mechanical system and calculate the
oscillating variance of position of the wave packet.
The variance of position operator for the Dirac Hamiltonian was calculated by
Barut and Malin~\cite{Barut1968} who found it to be the reminiscence of ZB of electrons in a vacuum.
The present authors analyzed in Ref.~\cite{Rusin07b} the variance of position operator
in bilayer graphene and found its oscillating character with the frequency equal to that of ZB.

One should finally remark that the attempts are constantly made in the literature to overcome
the above mentioned difficulties in the interpretation of position operator in KG equation.
In particular, Mostafazadeh~\cite{Mostafazadeh2004} proposed a redefinition of the
scalar product of solutions to KGE which allows one to obtain positively defined
probability distribution of position. Semenov {\it et al.}~\cite{Semenov2008}
proposed to limit the allowed solutions of KG equation to those having
positive-definite probability distributions. They showed that the physical
solutions of KGE fulfill this criterion. If the above attempts are accepted
one could analyze ZB of the position operator for KG particles, see Eq.~(\ref{H_rz_11q}).

\section{Summary \label{Sec_Sum}}

We considered the trembling motion (Zitterbewegung) of relativistic spin-zero particles
in absence of fields and in the presence of a magnetic field using the Klein-Gordon equation.
We aimed to describe physical observables (currents and velocities)
calculating quantities averaged with the use of Gaussian wave packets.
Surprisingly, the calculated particle
velocities can exceed the velocity of light for sufficiently large momenta indicating that
KGE does not posses an automatic restriction of relativity. We showed that the trembling
motion has a decaying character resulting from an interference of positive and negative energy
sub-packets moving with different velocities. In the presence of a magnetic field there
exist many interband frequencies that contribute to Zitterbewegung.
On the other hand, in the limit of non-relativistic energies the interband ZB components
vanish while the intraband components reduce to the cyclotron motion with a single
frequency. The trembling motion was simulated using the classical system obeying
the Klein-Gordon equation -- a stretched string attached to a rubber sheet. The calculated
variance of position of the sting shaped initially as a Gaussian packet
exhibits oscillations corresponding to Zitterbewegung with the correct frequency.

\appendix
\section{} \label{App_BJ}
In this Appendix we calculate an exact time dependence of current operators for a KG
particle in a magnetic field. We define the creation and annihilation operators
\begin{equation}
 \begin{array}{ccc} \label{BJ_aap_def}
     \ha  &=& (\xi+ \partial/\partial \xi)/\sqrt{2}, \\
     \hap &=& (\xi -\partial/\partial \xi)/\sqrt{2}, \end{array}
\end{equation}
and rewrite Eq.~(\ref{B_H}) in the form
\begin{equation} \label{BJ_H}
 \hH =\hT\left[\hbar\omega_c\left(\hap\ha +\frac{1}{2}\right) + \frac{\hbar^2k_z^2}{2m} \right]
  +\tau_3mc^2,
\end{equation}
where~$\omega_c=qB/m$ is the cyclotron frequency and~$\hT=(\tau_3+i\tau_2)$.
The current density is
\begin{eqnarray} \label{BJ_jj}
{\bm j} &=& \frac{\hbar}{2im}\left[\psi^{\dagger}\tau_3\hT{\bm \nabla}\psi -
            ({\bm \nabla}\psi^{\dagger})\tau_3\hT\psi \right] \nonumber \\
        &&- \frac{e}{mc}{\bm A}\psi^{\dagger}\tau_3\hT\psi,
\end{eqnarray}
and the average current is~$\langle{\bm j}\rangle=\int {\bm j}d^3{\bm r}$.
We introduce the {\it current operator}~$\hat{\bm J}$ in such way that
for~${\bm j}$ given in Eq.~(\ref{BJ_jj}) there is
\begin{equation} \label{BJ_Op}
 \langle \psi|\tau_3\hat{\bm J}|\psi\rangle = \int {\bm j}d^3{\bm r}.
\end{equation}
Note the presence of~$\tau_3$ in the matrix element.
In the asymmetric gauge one has
\begin{eqnarray} \label{BJ_jxAv0}
\langle j_x \rangle &=&
   -\frac{i\hbar}{2m} \int \left( \psi^{\dagger} \tau_3\hT \frac{\partial\psi}{\partial x}
   -\frac{\partial\psi^{\dagger}}{\partial x} \tau_3\hT \psi \right) d^3{\bm r} \nonumber \\
&& -\frac{qB}{mc}\int \left(\psi^{\dagger} \tau_3\hT y \psi \right) d^3{\bm r}, \ \\
                  \label{BJ_jyAv0}
\langle j_y \rangle &=& -\frac{i\hbar}{2m} \int \left( \psi^{\dagger} \tau_3\hT \frac{\partial\psi}{\partial y} -
     \frac{\partial\psi^{\dagger}}{\partial y} \tau_3\hT \psi \right) d^3{\bm r}.
\end{eqnarray}
Below we assume the function~$\psi$ to be Gaussian-like. In that case we may
simplify the above expressions for the average current by integrating by parts
the terms including derivatives of~$\psi^{\dagger}$
\begin{eqnarray} \label{BJ_jxAv}
\langle j_x \rangle &=&
  -\frac{i\hbar}{m} \int \left(\psi^{\dagger} \tau_3\hT \frac{\partial\psi}{\partial x}\right) d^3{\bm r}
      - \frac{qB}{mc}\int \left(\psi^{\dagger} \tau_3\hT y \psi \right) d^3{\bm r}, \ \\
     \label{BJ_jyAv}
\langle j_y \rangle &=& -\frac{i\hbar}{m} \int \left( \psi^{\dagger} \tau_3\hT \frac{\partial\psi}{\partial y}
          \right) d^3{\bm r},
\end{eqnarray}
so the components of the current operator are
\begin{eqnarray} \label{BJ_hjx}
\hat{J}_x &=& -\frac{i\hbar}{m} \hT \frac{\partial}{\partial x} -\frac{qB}{mc} \hT \hat{y}, \\
               \label{BJ_hjy}
\hat{J}_y &=& -\frac{i\hbar}{m} \hT \frac{\partial}{\partial y}.
\end{eqnarray}
In the Heisenberg picture the time-dependent current operator is
\begin{equation} \label{BJ_Jt}
\hat{\bm J}(t) = e^{i\hH t/\hbar} \hat{\bm J}(0) e^{-i\hH t/\hbar},
\end{equation}
where~$\hH$ is given in Eq.~(\ref{BJ_H}).
Our task is to calculate the time evolution of the current operator~$\hat{J}_x(t)$ and~$\hat{J}_y(t)$.
By averaging these operators over the state~$\psi$,
as shown in Eq.~(\ref{BJ_Op}), one obtains the time-dependent
charge current corresponding to~$\psi$.

It is convenient rewrite current operators in Eqs.~(\ref{BJ_hjx})
and~(\ref{BJ_hjy}) in the form
\begin{eqnarray} \label{BJ_hjx_a}
 \hat{J}_x &=& -\frac{i\hbar}{m}\hP
               -\frac{qB}{\sqrt{2}mc} \left(\hJ + \hJp\right) , \\
               \label{BJ_hjy_a}
 \hat{J}_y &=& -\frac{i\hbar}{\sqrt{2}m}\left(\hJ - \hJp\right),
\end{eqnarray}
where we introduce three auxiliary operators:
\begin{eqnarray} \
 \hP &=& (\tau_3 + i\tau_2) \frac{\partial}{\partial x} \equiv \hT \frac{\partial}{\partial x}, \\
 \label{BJ_def_hJ}
 \hJ &=& (\tau_3 + i\tau_2) \ha \equiv \hT \ha, \\
 \label{BJ_def_hJp}
 \hJp &=& (\tau_3 + i\tau_2) \hap \equiv \hT \hap.
\end{eqnarray}
We calculate the time dependence of~$\hJ$,~$\hJp$ and~$\hP$ in a way similar to
that described in Ref.~\cite{Rusin2010}. Consider first the operator~$\hP$.
From the equation of motion~$\hP_t=(i/\hbar)[\hH,\hP]$ one has
\begin{equation}
\hP_t=\frac{imc^2}{\hbar}[\tau_3,\hP]=2i\omega_0 \tau_1 \frac{\partial}{\partial x},
\end{equation}
where we used~$\hT^2=0$. Since~$\{\hH,\hP_t \}=0$,
there is~$[\hH,\hP_t]=2\hH\hP_t-\{\hH,\hP_t \} =2\hH\hP_t$, and one obtains
\begin{equation}
\hP_{tt}=\frac{2i}{\hbar}\hH \hP_t.
\end{equation}
We solve this equation for~$\hP_t$ and then integrate the solution over time
\begin{equation}
 \hP(t) =\frac{\hbar}{2i\hH}e^{2i\hH t/\hbar} \hP_t(0) + \hC,
\end{equation}
where~$\hC$ is a constant of integration. Applying the initial
conditions:~$\hP(0)=\hT(\partial/\partial x)$,~$\hP_t(0)=2i\omega_0 \tau_1 (\partial/\partial x)$,
and using the identity~$\hH^{-1}=\hH/E^2$ we have
\begin{equation} \label{BJ_Pt}
 \hP(t) = \hT\frac{\partial}{\partial x} + \frac{\hbar\omega_0\hH}{E^2}\left(e^{2i\hH t/\hbar}-1 \right)
 \tau_1 \frac{\partial}{\partial x}.
\end{equation}
It is seen that~$\hP(t)$ in Eq.~(\ref{BJ_Pt}) satisfies the initial conditions
for~$\hP(0)$ and~$\hP_t(0)$. The form of~$\hP(t)$ given above resembles results obtained
for the position operator in the field-free case by Fuda and Furlani~\cite{Fuda1982}.

Now we turn to the operators~$\hJ$ and~$\hJp$. From Eqs.~(\ref{BJ_def_hJ})
and~(\ref{BJ_def_hJp}) one has
\begin{eqnarray}
\label{BJ_hJt}  \hJ_t  &=& 2i\omega_0 \tau_1\ha, \\
\label{BJ_hJpt} \hJp_t &=& 2i\omega_0 \tau_1\hap,
\end{eqnarray}
where~$\omega_0=mc^2/\hbar$. Using~$[\ha,\hap]=1$ one obtains
\begin{eqnarray}
\label{BJ_aJt}  \{\hH,\hJ_t\}  &=& -2i \omega_0\hbar\omega_c \hJ,\\
\label{BJ_aJpt} \{\hH,\hJp_t\} &=& +2i\omega_0\hbar\omega_c \hJp.
\end{eqnarray}
Upon applying the identities
\begin{eqnarray}
[\hH, \hJ_t ] &=& +2\hH\hJ_t\ \   - \{\hH,\hJ_t\}, \\ \protect
[\hH, \hJp_t] &=& -2\hJp_t\hH +\{\hH,\hJp_t \},
\end{eqnarray}
we get
\begin{eqnarray} 
\label{BJ_hJtt}  \hJ_{tt}  &=& +(2i/\hbar)\hH  \hJ_t\ \ - 2\omega_0\omega_c\hJ,\\
\label{BJ_hJptt} \hJp_{tt} &=& -(2i/\hbar)\hJp_t \hH    - 2\omega_0\omega_c\hJp.
\end{eqnarray}
In Eqs.~(\ref{BJ_hJtt}) and~(\ref{BJ_hJptt}) we eliminate terms with the first derivatives
using the substitutions~$\hJ = \exp(+i\hH t/\hbar)\hB$ and~$\hJp = \hBp\exp(-i\hH t/\hbar)$,
respectively. This gives
\begin{eqnarray}
 \label{BJ_Btt}  \hB_{tt}  &=& -(\hO^2 + 2\omega_c\omega_0)\hB, \\
 \label{BJ_Bptt} \hBp_{tt} &=& -\hBp(\hO^2 + 2\omega_c\omega_0),
\end{eqnarray}
where~$\hO = \hH/\hbar$. In the above equations the operator
\begin{equation} \label{BJ_def_M}
 \hM^2 = \hO^2 + 2\omega_c\omega_0
\end{equation}
stands on the left-hand side of~$\hB$, but on the right-hand
side of~$\hBp$. Solutions to Eqs.~(\ref{BJ_Btt}) and~(\ref{BJ_Bptt}) are
\begin{eqnarray}
 \hB  &=& e^{-i\hM t}\hC_1  + e^{i\hM t} \hC_2, \\
 \hBp &=& \hCp_1e^{-i\hM t} + \hCp_2 e^{i\hM t},
\end{eqnarray}
where~$\hM =+\sqrt{\hM^2}$ is the positive root of~$\hM^2$.
Both~$\hC_1$ and~$\hC_2$ and their complex conjugates are time-independent operators.

Using the initial conditions:~$\hB(0)=\hJ(0)=\hT\ha$ and
$\hB_t(0)=\hJ_t(0)=+2i\omega_0 \tau_1\ha$ [see Eq.~(\ref{BJ_hJt})] and
similar expressions for~$\hBp(0)$ and~$\hBp_t(0)$, we find
that~$\hJ(t)=\hJ_1(t)+\hJ_2(t)$, where
\begin{eqnarray}
\label{BJ_hJ1} \hJ_1(t) &=& \frac{1}{2}e^{i\hO t}e^{-i\hM t}\left[\hJ(0) + \hM^{-1}\hJ(0)\hO \right], \\
\label{BJ_hJ2} \hJ_2(t) &=& \frac{1}{2}e^{i\hO t}e^{+i\hM t}\left[\hJ(0) - \hM^{-1}\hJ(0)\hO \right].
\end{eqnarray}
Similarly, one can express~$\hJp(t)=\hJp_1(t)+\hJp_2(t)$, where
\begin{eqnarray}
\label{BJ_hJp1} \hJp_1(t) &=& \frac{1}{2}\left[\hJp(0) + \hO\hJp(0)\hM^{-1}\right]e^{+i\hM t}e^{-i\hO t},\ \ \\
\label{BJ_hJp2} \hJp_2(t) &=& \frac{1}{2}\left[\hJp(0) - \hO\hJp(0)\hM^{-1}\right]e^{-i\hM t}e^{-i\hO t}.\ \
\end{eqnarray}
The results are given in terms of the operators~$\hO$ and~$\hM$.
To finalize the description, one needs
to specify the physical sense of functions appearing in
Eqs.~(\ref{BJ_hJ1}) -~(\ref{BJ_hJp2}).

For a reasonable function~$f(\hat{D})$ of an operator~$\hat{D}$,
its eigenenergies~$\lambda_{\rm d}$ and its eigenstates~$|{\rm d}\rangle$,
there exists the following
relationship:~$f(\hat{D})|{\rm d}\rangle = f(\lambda_{\rm d})|{\rm d}\rangle$,
provided that~$f(\lambda_{\rm d})$ exists. To find the meanings of the
operators~$M^{-1}$ and~$e^{\pm i\hM T}$ we express them as
functions of the operator~$\hM^2=\hH^2/\hbar^2+2\omega_0\omega_c$, see Eq.~(\ref{BJ_def_M}).
From the definition of~$\hM^2$ it follows that its
eigenstates are equal to the eigenstates~$|{\rm n}\rangle$ of~$\hH$.
The eigenvalues~$\lambda_{\rm n}^2$ of the operator~$\hM^2$
are~$\lambda_{n,k_z}^2=E_{n+1,k_z}^2$ and we obtain
\begin{eqnarray}
\label{BJ_hM1} \hM^{\pm 1}|{\rm n}\rangle &=& (\hM^2)^{\pm 1/2}|{\rm n}\rangle =
     \eta E_{n+1,k_z}^{\pm 1} |{\rm n}\rangle, \\
\label{BJ_ehM1} e^{\pm i\hM t}|{\rm n}\rangle &=& e^{\pm i(\hM^2)^{1/2} t}|{\rm n}\rangle =
    e^{\pm i\eta E_{n+1,k_z}} |{\rm n}\rangle,
\end{eqnarray}
where~$\eta=+1$ or~$\eta=-1$. As seen from Eqs.~(\ref{BJ_hJ1}) -~(\ref{BJ_hJp2}), the
sums~$\hJ_1(t)+\hJ_2(t)$ and~$\hJp_1(t)+\hJp_2(t)$ do not depend on the sign of~$\eta$,
so we select~$\eta=+1$.

Finally we show that the matrix elements of the operator~$\hJ(t)=\hJ_1(t)+\hJ_2(t)$
are equal to the matrix elements of the current operator~$\hJ_H(t)=e^{i\hO t}\hJ(0)e^{-i\hO t}$
in the Heisenberg picture. The operator~$\hJ$ is proportional to the annihilation
operator~$\ha$ whose non-vanishing matrix elements
are~$\langle n'|\ha|n\rangle =\sqrt{n+1}\delta_{n',n+1}$, so
we select two eigenstates of KG
Hamiltonian~$|{\rm n}\rangle=|n,s\rangle$ and~$|{\rm n}'\rangle=|n+1,z\rangle$,
see Eq.~(\ref{B_Psi}). Here we omitted quantum numbers~$k_x$ and~$k_z$.
For~$\hJ_H(t)$ one has
\begin{equation} \label{BJ_JHnn}
\langle {\rm n}|\tau_3\hJ_H(t)|{\rm n}'\rangle = e^{is\omega_nt}e^{-iz\omega_{n+1}t}
 \hJ(0)_{\rm n n'},
\end{equation}
where we define~$\hJ(0)_{\rm n n'}=\langle {\rm n}|\tau_3\hJ(0)|{\rm n}'\rangle$. To calculate
the matrix elements of~$\hJ_1(t)$ we use Eqs.~(\ref{BJ_hM1}) -~(\ref{BJ_ehM1}) and obtain
\begin{equation}
\langle {\rm n}|\hM^{-1}\hJ(0)\hO|{\rm n}'\rangle =
    \frac{\hbar}{E_{n+1}}\hJ(0)_{\rm n n'}\frac{z E_{n+1}}{\hbar}
  = z \hJ(0)_{\rm n n'}
\end{equation}
which finally gives
\begin{eqnarray}
\label{BJ_J1nn} \langle {\rm n}|\hJ_1(t)|{\rm n}'\rangle
       &=&\frac{1+z}{2}\hJ(0)_{\rm n n'}e^{is\omega_nt}e^{-i\omega_{n+1}t}, \\
\label{BJ_J2nn} \langle {\rm n}|\hJ_2(t)|{\rm n}'\rangle
       &=&\frac{1-z}{2}\hJ(0)_{\rm n n'}e^{is\omega_nt}e^{+i\omega_{n+1}t}.
\end{eqnarray}
The matrix elements of~$\hJ_1(t)$ are nonzero for~$z=+1$ only, while the matrix
elements of~$\hJ_2(t)$ are nonzero for~$z=-1$ only. Comparing Eqs.~(\ref{BJ_J1nn}) and~(\ref{BJ_J2nn})
with Eq.~(\ref{BJ_JHnn}) we see that for each of four combinations of~$s=\pm 1$ and~$z=\pm 1$
the matrix elements of~$\hJ_H(t)$ are equal to the matrix elements of~$\hJ(t)=\hJ_1(t)+\hJ_2(t)$,
which is what we wanted to show. Calculations for~$\hJp(t)$ are similar to those for~$\hJ(t)$.
The compact equations (\ref{BJ_hJ1}) -~(\ref{BJ_hJp2}) are our final results
for the time dependence of~$\hJ(t)$ and~$\hJp(t)$
operators. These equations are exact and they are
quite fundamental for relativistic spin-0 particles in a magnetic field.
If we calculate average currents of Eqs.~(\ref{BJ_hjx_a}) and~(\ref{BJ_hjy_a})
with the use of expressions~(\ref{BJ_J1nn}) -~(\ref{BJ_J2nn})
and the wave packet~(\ref{P_wr}), one obtains
results corresponding to the velocities given in Section~\ref{Sec_Field}.

\section{} \label{App_Light}

In this Appendix we analyze in more detail the relation of the particle velocity
to the speed of light. We consider~$(1,1)$ component of the velocity operator for
a KG particle given in Eq.~(\ref{H_vz_11}).
For the wave packet~$\langle{\bm r}|w\rangle= w({\bm r})(1,0)^T$ with one nonzero component
the average velocity is given by the average of~$(\hvz)_{11}(t)$ over the function~$w({\bm r})$.
The unexpected feature of operator~$(\hvz)_{11}(t)$ is that
for large~${\bm p}$ this velocity can exceed the speed of light~$c$.

There are two possible ways to overcome this problem. We can {\it additionally}
assume that~$|p|\le mc$, which ensures that the velocity~$(\hvz)_{11}(t)$ does not exceed~$c$.
This condition is equivalent to~$|q|\le 1$ in the text, see Eq.~(\ref{H_vz_11}).
Alternatively, one can take the initial wave packet~$w({\bm r})$ which does not contain
components with~$|{\bm p}|> mc$. Then the Gaussian packet in Eq.~(\ref{P_wk}) must be replaced by a
non-Gaussian packet~$w'({\bm r})$ of the form
\begin{equation} \label{App_Light_wk}
 \langle {\bm k}|w'\rangle=(2d\sqrt{\pi})^{3/2}\exp[-d^2({\bm k}-{\bm k}_0)^2/2]\Theta(\lambda_c-|{\bm k}|),
\end{equation}
where~$\Theta(\xi)$ is the step function.

For the Dirac Hamiltonian~$\hH_D= c\sum_{j}\hat{\alpha}_j \hat{p}_j + mc^2\hat{\beta}$,
the situation is different. Expanding~$e^{i\hH_Dt/\hbar}$ in a power series one obtains an expression
analogous to~$e^{i\hH t/\hbar}$ given in Eq.~(\ref{H_eiHt}). After some algebra
we find
\begin{equation} \label{App_Light_Dirac}
(\hvz)_{11}^D(t)=\frac{mc^2p_z}{m^2c^2+p^2} \left[1 - \cos(2Et/\hbar)\right].
\end{equation}
In contrast to the KG case, the velocity operator given in Eq.~(\ref{App_Light_Dirac})
has correct relativistic behavior for all values of~${\bm p}$.
In Eq.~(\ref{App_Light_Dirac}) the expression in square brackets oscillates between zero and two.
The factor~$v^D({\bm p}) = mc^2p_z/(m^2c^2+p^2)$ tends to zero for both
large and small values of~${\bm p}$. Its maximum is at~${\bm p}^{max}=(0,0,mc)$
for which one obtains
\begin{equation} \label{App_Light_DiracMax}
(\hvz)_{11}^D(t)=\frac{c}{2} \left[1 - \cos(2\sqrt{2}\omega_0)\right].
\end{equation}
The above velocity never exceeds the speed of light.
Therefore, when calculating the average velocity of the wave packet for the Dirac
Hamiltonian, there is no need for an artificial truncation of the high
momentum components of the wave packet, as proposed in Eq.~(\ref{App_Light_wk}) for a KG particle.

\section{} \label{App_Ident}
We prove here some identities appearing in the previous sections. We begin with
the identity in Eq.~(\ref{Bv_one}). Closing Eq.~(\ref{Bv_one}) with the use of
states~$\langle {\bm r}|$ and~$|{\bm r}'\rangle$, employing
Eq.~(\ref{B_Psi}) and writing explicitly the summations and integrations over the quantum numbers
we obtain
\begin{eqnarray} \label{App_Ident_1}
\delta_{{\bm r},{\bm r}'}&=&\sum_{\rm n}\langle {\bm r}|{\rm n}\rangle\langle{\rm n}|{\bm r}'\rangle
 s_{\rm n} \tau_3 = \frac{1}{16\pi^2} \sum_{n=0}^{\infty}\phi_n(\xi)\phi_n(\xi')
  \nonumber \\
 & \times & \int_{-\infty}^{\infty} e^{ik_x(x-x')}dk_x
          \int_{-\infty}^{\infty} e^{ik_z(z-z')}dk_z
 \nonumber \\
 & \times & \sum_{s=\pm 1}\left(\begin{array}{c} \mu^+ \\ \mu^- \end{array}\right)
          \left(\mu^+ ,\mu^-\right)s \tau_3,
\end{eqnarray}
where~$\mu^{\pm} \equiv \mu_{n,k_z,s}^{\pm}$
In the above equation the summation over~$n$ gives~$\delta_{\xi,\xi'}$, the product of the two
integrals is~$4\pi^2\delta_{x,x'}\delta_{z,z'}$, so the product of the three terms
equals~$4\pi^2\delta_{{\bm r},{\bm r}'}$. Taking the explicit form of~$\mu^{\pm}=\nu\pm s/\nu$
where~$\nu=\sqrt{mc^2/E_{n,k_z}}$, we obtain for the last line of Eq.~(\ref{App_Ident_1})
\begin{equation}
\sum_{s=\pm 1}
\left(\begin{array}{cc}
    s(\nu+s/\nu)^2   & s(\nu^2-1/\nu^2) \\
    s(\nu^2-1/\nu^2) & s(\nu-s/\nu)^2 \end{array} \right)
    \left(\begin{array}{cc} 1 & 0 \\ 0 & -1 \end{array} \right) = 4.
\end{equation}
Collecting all numerical factors we see that the right hand side of
Eq.~(\ref{App_Ident_1}) equals~$\delta_{{\bm r},{\bm r}'}$.

Next we prove the identity used in the derivation of Eqs.~(\ref{Av_tauH}) and~(\ref{Bv_tauH}).
Let~$\hcO$ be any operator for which~$\hcO=\tau_3\hcO^{\dagger}\tau_3$, where
dagger signifies the Hermitian conjugate. We want to show that
\begin{equation} \tau_3e^{\hcO}=e^{\hcO^{\dagger}}\tau_3. \end{equation}
To this end we expand the exponents and get
\begin{equation} \label{App_Ident_Ex}
\tau_3\left[1 + \frac{\hcO}{1!} +\frac{\hcO^2}{2!} \ldots \right] =
      \left[1 + \frac{\hcO^{\dagger} }{1!}+\frac{\hcO^{\dagger 2} }{2!}\ldots \right]\tau_3.
\end{equation}
Since~$\hcO=\tau_3\hcO^{\dagger}\tau_3$, there is~$\hcO^{\dagger}=\tau_3\hcO\tau_3$.
Then for~$n\ge 0$ there is~$\hcO^{\dagger n}=\tau_3\hcO^n\tau_3$
and we obtain for the RHS of Eq.~(\ref{App_Ident_Ex})
\begin{eqnarray*}
 \left[1 + \frac{\hcO^{\dagger} }{1!}+\frac{\hcO^{\dagger 2} }{2!}\ldots \right]\tau_3 =
 \left[1 + \tau_3\frac{\hcO }{1!}\tau_3+ \tau_3\frac{\hcO^{2} }{2!}\tau_3\ldots \right]\tau_3
 \nonumber \\
  = \tau_3 \left[1 +\frac{\hcO }{1!}+ \frac{\hcO^{2} }{2!}\ldots \right] =\tau_3 e^{\hcO},
\end{eqnarray*}
which is the desired result.

\section{} \label{App_Gauss}
In this appendix we quote for completeness all formulas necessary for
a calculation of coefficients~$U_{m,n}$ in Eqs.~(\ref{Bv_vyt}) -~(\ref{Bv_vzt}).
Here we assume the initial wave vector in the form~${\bm k}_{0}=(k_{0x},0,k_{0z})$.
Using the definitions of~$g_{xy}(k_x,y)$,~$F_n(k_x)$ and~$U_{m,n}$,
we obtain (see Ref.~\cite{Rusin2010})

\begin{equation} \label{App_Gauss_gxy}
g_{xy}(k_x,y) = \sqrt{\frac{d_x}{\pi d_y}}e^{-\frac{1}{2}d_x^2(k_x-k_{0x})^2} e^{-\frac{y^2}{2d_y^2}},
\end{equation}
and
\begin{equation} \label{App_Gauss_Fn}
 F_n(k_x) = \frac{A_n\sqrt{L d_x}}{\sqrt{2\pi d_y}C_n} e^{-\frac{1}{2}d_x^2(k_x-k_{0x})^2}
            e^{-\frac{1}{2}k_x^2D^2}\ {\rm H}_n(-k_xc),
\end{equation}
where~$D=L^2/\sqrt{L^2+d_y^2}$,~$c=L^3/\sqrt{L^4-d_y^4}$, and
\begin{equation}
 A_n=\frac{\sqrt{2\pi}d_y}{\sqrt{L^2+d_y^2}}\left(\frac{L^2-d_y^2}{L^2+d_y^2}\right)^{n/2},
\end{equation}
\begin{eqnarray} \label{App_Gauss_Umn}
U_{m,n}= \frac{A_m^* A_nLQd_x \sqrt{\pi}\ e^{-W^2}}{\pi C_mC_n d_y}
    \sum_{l=0}^{\min\{m,n\}}\!\!\!\! 2^l l! \! \left(\begin{array}{c} m \\ l\!\! \end{array}\right) \!\!\!
     \left(\begin{array}{c} n \\ l\!\! \end{array}\right)
   && \nonumber \\
  \times (\left(1-(cQ)^2\right)^{(m+n-2l)/2}
    {\rm H}_{m+n-2l} \left(\frac{-cQY}{\sqrt{1-(cQ)^2}}\right), \ \ \ \ \ &&
\end{eqnarray}
in which~$Q = 1/\sqrt{d_x^2+D^2}$,~$W= d_xDQk_{0x}$, and~$Y=d_x^2k_{0x}Q$.
For the special case of~$d_y=L$, the formula for~$U_{m,n}$ is much simpler:
\begin{eqnarray} \label{App_Gauss_Umn_Ldy}
U_{m,n} &=& 2\frac{\sqrt{\pi}\ (-i)^{m+n}\ d_x}{C_mC_nL}
   \left(\frac{L}{2P} \right)^{m+n+1} \times \nonumber \\
 && \exp\left(-\frac{d_x^2k_{0x}^2L^2}{2P^2}\right)
     {\rm H}_{m+n}\left(\frac{-id_x^2k_{0x}}{P}\right), \ \ \
\end{eqnarray}
where~$P=\sqrt{d_x^2+\frac{1}{2}L^2}$. In the above expressions
the coefficients~$U_{m,n}$ are real numbers and they are symmetric
in~$m,n$ indices. For further discussion of
of~$U_{m,n}$ see Refs.~\cite{Rusin2010,Rusin2008a}

\end{document}